\documentclass[acmsmall]{acmart}

\usepackage[english]{babel}
\usepackage{natbib}

\usepackage{amsmath}
\usepackage{amsfonts}
\usepackage{bbm}
\usepackage{bm}
\usepackage{mathtools}
\usepackage{scalerel} 

\usepackage{booktabs} 
\usepackage[online,flushleft]{threeparttable}
\usepackage{multirow}
\usepackage{subcaption}
\usepackage{adjustbox}
\usepackage{graphicx}
\usepackage{ulem}

\usepackage[justification=justified]{caption}

\usepackage[inline]{enumitem}
\usepackage[capitalize]{cleveref}

\usepackage{color}
\usepackage{tikz}
\usetikzlibrary{shapes, positioning, fit, calc}
\usetikzlibrary{backgrounds}
\usetikzlibrary{arrows.meta}
\usetikzlibrary{shapes.arrows, shapes.geometric}
\usetikzlibrary{matrix}

\usepackage{pgfplots}
\usepgfplotslibrary{groupplots}
\pgfplotsset{compat=1.14}

\usepackage{soul}
\setstcolor{red}
\usepackage{sistyle}
\SIthousandsep{,}

\AtBeginDocument{%
  \providecommand\BibTeX{{%
    \normalfont B\kern-0.5em{\scshape i\kern-0.25em b}\kern-0.8em\TeX}}}

\usepackage{xcolor}

\definecolor{oyster_pink}{RGB}{238,206,205} 
\definecolor{coral_candy}{RGB}{242,208,205} 
\definecolor{baby_pink}{RGB}{246, 194, 192}
\definecolor{oyster_pink}{RGB}{238,206,205} 
\definecolor{NY_pink}{RGB}{228,136,113} 
\definecolor{petite_orchid}{RGB}{223, 157, 155}
\definecolor{carmine_pink}{RGB}{231, 76, 60}
\definecolor{deep_carmine_pink}{RGB}{236, 50, 67}
\definecolor{apricot}{RGB}{241,140,122}

\definecolor{milan}{RGB}{255, 254, 169}
\definecolor{casablanca}{RGB}{244, 178, 84}
\definecolor{texas}{RGB}{245, 232, 123}
\definecolor{maize}{RGB}{249, 212, 156}

\definecolor{double_pearl_lusta}{RGB}{253, 242, 208}

\definecolor{oasis}{RGB}{253, 242, 208}
\definecolor{linen}{RGB}{251, 239, 227}

\definecolor{zanah}{RGB}{220, 233, 213}
\definecolor{frostee}{RGB}{217, 231, 214}
\definecolor{norway}{RGB}{158, 194, 132}
\definecolor{malibu}{RGB}{110, 180, 240}

\definecolor{link_water}{RGB}{221, 232, 250}

\definecolor{spring_leaves}{RGB}{46, 83, 117}

\definecolor{venice_blue}{RGB}{87, 135, 105}
\definecolor{boston_blue}{RGB}{68, 147, 161}

\definecolor{napa}{RGB}{163, 154, 137}

\definecolor{mexican_red}{RGB}{170, 41, 37}
\definecolor{valencia}{RGB}{214, 87, 70}

\definecolor{riptide}{RGB}{141,211,199}
\definecolor{pale_prim}{RGB}{255,255,179}
\definecolor{lavender_gray}{RGB}{190,186,218}
\definecolor{salmon}{RGB}{242,131,107}
\definecolor{seagull}{RGB}{128,177,211}
\definecolor{rajah}{RGB}{253,180,98}
\definecolor{yellow_green}{RGB}{198,222,119}
\definecolor{classic_rose}{RGB}{252,205,229}
\definecolor{feijoa}{RGB}{178,223,138}

\definecolor{cruise}{RGB}{179,226,205}

\definecolor{periwinkle}{RGB}{203,213,232}
\definecolor{snow_flurry}{RGB}{230,245,201}
\definecolor{buttermilk}{RGB}{255,242,174}

\definecolor{sundown}{RGB}{249, 180, 181}
\definecolor{spindle}{RGB}{179,205,227}
\definecolor{tea_green}{RGB}{204,235,197}
\definecolor{languid_lavender}{RGB}{222,203,228}
\definecolor{champagne}{RGB}{254,217,166}
\definecolor{cream}{RGB}{255,255,204}

\definecolor{monte_carlo}{RGB}{135,204,194}
\definecolor{melon}{RGB}{254,191,181}
\definecolor{granny_smith_apple}{RGB}{150,214,150}
\definecolor{watusi}{RGB}{254,221,207}
\definecolor{see_green}{RGB}{161,228,195}

\definecolor{moss_green}{RGB}{170,216,176}
\definecolor{opal}{RGB}{164,207,190}

\definecolor{pale_turquoise}{RGB}{172,240,242}
\definecolor{Madang}{RGB}{190,235,159}
\definecolor{pixie_green}{RGB}{183,214,170}
\definecolor{coral_andy}{RGB}{243,204,205}
\definecolor{manhattan}{RGB}{226,180,125}
\definecolor{quartz}{RGB}{219,223,238}
\definecolor{spring_sun}{RGB}{242,243,195}
\definecolor{dairy_cream}{RGB}{254,226,189}
\definecolor{surf_crest}{RGB}{205,230,208}
\definecolor{french_pass}{RGB}{195,232,246}
\definecolor{cosmos}{RGB}{248,209,210}
\definecolor{portafino}{RGB}{245,237,160}
\definecolor{sail}{RGB}{163,205,235}
\definecolor{hint_green}{RGB}{226,246,209}

\definecolor{jet_stream}{RGB}{188, 214, 210}

\definecolor{azalea}{RGB}{251, 196, 196}
\definecolor{wewak}{RGB}{244, 143, 150}
\definecolor{bittersweet}{RGB}{255,111,105}
\definecolor{sunset_orange}{RGB}{242,89,75}
\definecolor{light_coral}{RGB}{244, 127, 123}
\definecolor{carnation}{RGB}{245, 80, 86}
\definecolor{flamingo}{RGB}{237, 88, 85}

\definecolor{fire_engine_red}{RGB}{210,44,41}
\definecolor{amaranth}{RGB}{234,46,73}
\definecolor{ku_crimson}{RGB}{243, 0, 25}
\definecolor{fire_engine_red}{RGB}{206, 37, 51}
\definecolor{copper_rust}{RGB}{155, 64, 74}

\definecolor{chilean_fire}{RGB}{215, 87, 44}

\definecolor{japanese_laurel}{RGB}{53, 116, 40}

\definecolor{turmeric}{RGB}{211, 178, 76}
\definecolor{saffron}{RGB}{249,193,62}
\definecolor{my_sin}{RGB}{255, 176, 59}
\definecolor{tree_poppy}{RGB}{246, 154, 27}
\definecolor{jaffa}{RGB}{240, 131, 58}
\definecolor{crusta}{RGB}{254, 127, 44}
\definecolor{tahiti_gold}{RGB}{223, 102, 36}
\definecolor{outrageous_orange}{RGB}{255, 100, 45}
\definecolor{safety_orange}{RGB}{254, 106, 0}

\definecolor{turquoise}{RGB}{41,217,194}
\definecolor{puerto_rico}{RGB}{94, 194, 166}
\definecolor{mountain_meadow}{RGB}{0, 163, 136}
\definecolor{free_speech_aquamarine}{RGB}{0, 156, 114}
\definecolor{java}{RGB}{2,190,196}

\definecolor{matisse}{RGB}{25, 104, 167}
\definecolor{shakespeare}{RGB}{85, 154, 193}
\definecolor{mona_lisa}{RGB}{246,152,134}

\definecolor{bgc}{RGB}{245,245,245}
\definecolor{tuatara}{RGB}{67, 67, 67}
\definecolor{aluminum}{RGB}{153,153,153}
\definecolor{silver}{RGB}{191,191,191}
\definecolor{platinum}{RGB}{228,228,228}
\definecolor{mercury}{RGB}{230,230,230}
\definecolor{gallery}{RGB}{240,240,240}
\definecolor{athens_gray}{RGB}{236, 240, 241}
\definecolor{ship_gray}{RGB}{77,77,77}

\definecolor{early_dawn}{RGB}{252,243,218}
\definecolor{egg_shell}{RGB}{238, 234, 215}
\definecolor{midnight}{RGB}{0, 29, 50}
\definecolor{sundown}{RGB}{249, 180, 181}
\definecolor{sun_shade}{RGB}{255, 144, 68}
\definecolor{sushi}{RGB}{117, 168, 47}
\definecolor{tomato}{RGB}{255, 97, 56}
\definecolor{ice_cold}{RGB}{169,232,220}

\definecolor{jelly_bean}{RGB}{45, 126, 150}
\definecolor{celestial_blue}{RGB}{52, 152, 219}
\definecolor{curious_blue}{RGB}{41, 128, 185}
\definecolor{french_blue}{RGB}{0, 112, 182}
\definecolor{matisse}{RGB}{25, 104, 167}

\definecolor{biscay}{RGB}{44, 62, 80}

\definecolor{cosmic_latte}{RGB}{222, 247, 229}
\definecolor{chinook}{RGB}{163, 232, 178}
\definecolor{padua}{RGB}{121, 189, 143}
\definecolor{ocean_green}{RGB}{79, 176, 112}
\definecolor{pastel_green}{RGB}{107, 227, 135}
\definecolor{chateau_green}{RGB}{69, 191, 85}
\definecolor{RoyalBlue}{RGB}{69, 191, 85}
\definecolor{pigment_green}{RGB}{0, 175, 79}
\definecolor{fern}{RGB}{101,197,117}
\definecolor{killarney}{RGB}{56, 113, 66}
\definecolor{viridian}{RGB}{70, 137, 102}

\newcommand\blfootnote[1]{%
  \begingroup
  \renewcommand\thefootnote{}\footnote{#1}%
  \addtocounter{footnote}{-1}%
  \endgroup
}

\setcopyright{acmcopyright}
\acmJournal{TOIS}
\acmYear{2021} \acmVolume{1} \acmNumber{1} \acmArticle{1} \acmMonth{1} \acmPrice{15.00}\acmDOI{10.1145/3486250}

\usepackage{multirow}
\usepackage{tabularx}
\usepackage{makecell}

\renewcommand{\arraystretch}{1.2}

\begin{document}

\title{Semantic Models for the First-stage Retrieval: A Comprehensive Review}

\author{Jiafeng Guo, Yinqiong Cai, \lowercase{and} Yixing Fan$^{*}$}
\email{{guojiafeng,caiyinqiong18s,fanyixing}@ict.ac.cn}
\affiliation{
  \department{CAS Key Lab of Network Data Science and Technology}
  \institution{Institute of Computing Technology, Chinese Academy of Sciences; University of Chinese Academy of Sciences}
  \streetaddress{NO. 6 Kexueyuan South Road, Haidian District}
  \city{Beijing}
  \country{China}
  \postcode{100190}
}

\author{Fei Sun}
\email{ofey.sf@alibaba-inc.com}
\affiliation{
  \institution{Alibaba Group}
  \city{Beijing}
  \country{China}
  \postcode{100102}
}

\author{Ruqing Zhang, \lowercase{and} Xueqi Cheng$^{*}$}
\email{{zhangruqing,cxq}@ict.ac.cn}
\affiliation{
  \department{CAS Key Lab of Network Data Science and Technology}
  \institution{Institute of Computing Technology, Chinese Academy of Sciences; University of Chinese Academy of Sciences}
  \streetaddress{NO. 6 Kexueyuan South Road, Haidian District}
  \city{Beijing}
  \country{China}
  \postcode{100190}
}
\blfootnote{$^{*}$ Yixing Fan and Xueqi Cheng are the corresponding authors.}
\renewcommand{\authors}{Jiafeng Guo, Yinqiong Cai, Yixing Fan, Fei Sun, Ruqing Zhang, and Xueqi Cheng}
\renewcommand{\shortauthors}{Jiafeng Guo, Yinqiong Cai, Yixing Fan, Fei Sun, Ruqing Zhang, and Xueqi Cheng}

\begin{abstract}
Multi-stage ranking pipelines have been a practical solution in modern search systems, where the first-stage retrieval is to return a subset of candidate documents, and latter stages attempt to re-rank those candidates. Unlike re-ranking stages going through quick technique shifts during past decades, the first-stage retrieval has long been dominated by classical term-based models. Unfortunately, these models suffer from the vocabulary mismatch problem, which may block re-ranking stages from relevant documents at the very beginning. Therefore, it has been a long-term desire to build semantic models for the first-stage retrieval that can achieve high recall efficiently. Recently, we have witnessed an explosive growth of research interests on the first-stage semantic retrieval models. We believe it is the right time to survey current status, learn from existing methods, and gain some insights for future development. 
In this paper, we describe the current landscape of the first-stage retrieval models under a unified framework to clarify the connection between classical term-based retrieval methods, early semantic retrieval methods and neural semantic retrieval methods.
Moreover, we identify some open challenges and envision some future directions, with the hope of inspiring more researches on these important yet less investigated topics.
\end{abstract}



\maketitle

\section{Introduction}

\begin{figure}[!t]
\centering
\includegraphics[scale=1.0]{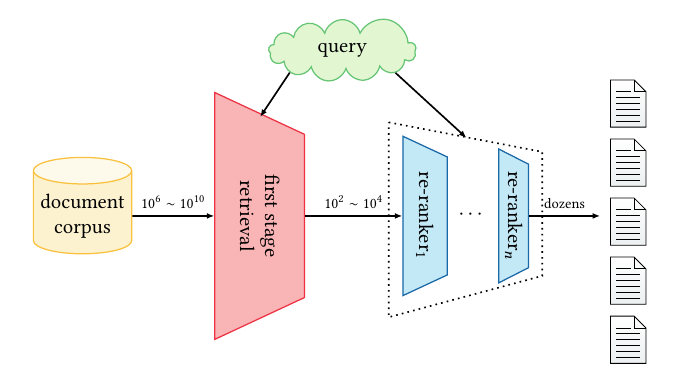}
\caption{The multi-stage architecture of modern information retrieval systems.} 
\label{fig:architecture}                                  
\end{figure}

Large-scale query-document retrieval is a key problem in search systems, e.g., Web search engines, which aims to return a set of relevant documents from a large document repository given a user query.
To balance the search efficiency and effectiveness, modern search systems typically employ a multi-stage ranking pipeline in practice, as shown in Figure \ref{fig:architecture}. 
The first-stage retrieval aims to return an initial set of candidate documents from a large repository by some cheaper ranking models assisted by some specially-designed indexing structures. Later, several re-ranking stages take more complex and effective ranking models to prune and improve the ranked document list output by the previous stage. Such a ``retrieval and re-ranking'' pipeline has been widely adopted in both academia~\cite{matveeva2006high, chen2017efficient} and industry~\cite{pedersen2010query, liu2017cascade} and achieved state-of-the-art results on multiple IR benchmarks \cite{voorhees2005overview, nguyen2016msmarco, dietz2017trec}.

Besides the pipeline architecture, to achieve a successful retrieval, it is generally recognized that the system needs to understand the query and the document well so that it can find relevant results to users' information needs. Therefore, semantic models are expected throughout the pipeline but with different requirements and goals at different stages. 
For the first-stage retrieval, the model aims to recall all potentially relevant documents from the whole collection. Thus, it is desired to build semantic models that can achieve high recall efficiently, i.e., to return a subset of documents that contain relevant documents as many as possible within a short time-span. For latter re-ranking stages, only a small number of documents are fed into the ranking model. As a result, semantic models used for re-ranking are allowed to employ more sophisticated architectures to achieve high precision, i.e., to put as many relevant documents as possible to top positions of the list.

During past decades, we have witnessed re-ranking stages going through quick technique shifts towards more and more powerful semantic models, from early probabilistic models \cite{robertson1976relevance, van1977theoretical, Robertson:Now:BM25}, learning to rank models \cite{li2011learning, liu2011learning}, to recent neural ranking models \cite{huang2013learning, guo2016deep, nogueira2019passage}. Specifically, with BERT-style pre-training tasks on cross-attention models, better contextualized representations and deeper interactions between query-document pairs have led to significant improvement on the re-ranking effectiveness~\cite{nogueira2019passage, nogueira2019multi}. 
However, these models are often very computationally expensive, which makes them unable to handle high-throughput incoming queries each with a large collection of candidate documents in the first-stage retrieval.

On the contrary, the first-stage retrieval has long been dominated by classical term-based models. Specifically, the discrete symbolic representation, i.e., bag-of-words (BoW) representation, is adopted for both queries and documents, and the inverted indexing technique is leveraged to manage large-scale documents. Term-based retrieval models such as BM25 (term matching + TF-IDF weights) are then applied for the first-stage retrieval. Apparently, such term-based models are very efficient due to the simple logic and powerful index. Meanwhile, they have also been demonstrated to achieve reasonable good recall performance in practice \cite{chen2017efficient, liu2017cascade}. However, there are still clear drawbacks with such term-based models: 
\begin{enumerate*}[label=(\arabic *)]
\item They may suffer from the \textsl{vocabulary mismatch} problem \cite{zhao2010term, furnas1987vocabulary} due to the independence assumption;
\item They may not well capture document semantics by ignoring term ordering information \cite{li2014semantic}.
\end{enumerate*}
Due to these limitations, term-based models may play as a ``blocker'' which prevents re-ranking models from relevant documents at the very beginning.
To resolve this problem, continuous efforts have been made during past decades, including query expansion~\cite{lesk1969word, qiu1993concept, lavrenko2017relevance, xu2017quary}, document expansion~\cite{efron2012improving, agirre2010document, liu2004cluster}, term dependency models~\cite{metzler2005markov, gao2004dependence, xu2010relevance}, topic models~\cite{deerwester1990indexing, wei2006lda}, translation models for IR~\cite{berger1999information, karimzadehgan2010estimation}, etc.
However, the research progress on the first-stage retrieval is relatively slow since most of these approaches are still within the discrete symbolic representation paradigm and inherit its limitations inevitably.

In recent years, along with the development of representation learning methods in information retrieval (IR), we have witnessed an explosive growth of research interests in the first-stage semantic retrieval models. Since 2013, the rise of word embedding technique~\cite{mikolov2013Distributed, pennington2014glove, bojanowski2017enriching} stimulates a large amount of work on exploiting it for the first-stage retrieval~\cite{clinchant2013aggregating, vulic2015monolingual, ganguly2015word}. Unlike the discrete symbolic representation, word embedding is a dense representation which may alleviate the vocabulary mismatch problem to some extent. After 2016, there is a surge of research interest in applying deep learning technique for the first-stage retrieval \cite{boytsov2016off, henderson2017efficient}. These approaches have been studied either to improve document representations within the conventional discrete symbolic representation paradigm~\cite{bai2020sparterm, dai2019context, nogueira2019document}, or directly form a new series of semantic retrieval models within the sparse/dense representation paradigm~\cite{zamani2018neural, jang2021uhd, gillick2018end, khattab2020colbert}. Since there has been a significant body of works created, we believe it is the right time to survey current status, learn from existing methods, and gain some insights for future development.

This survey focuses on semantic models for the first-stage retrieval of unstructured texts, referred to as \textit{semantic retrieval models} for short in the following sections. 
We describe the current landscape of the first-stage retrieval models under a unified framework to clarify the connection between classical term-based retrieval methods, early semantic retrieval methods and neural semantic retrieval methods. Specifically, we pay attention to recent neural semantic retrieval methods, summarizing them into three paradigms from the perspective of model architecture, namely sparse retrieval methods, dense retrieval methods and hybrid retrieval methods. We also refer to key topics about neural semantic retrieval models learning. Moreover, we discuss unresolved challenges and suggest potentially promising directions for future works.
It should be noted that: 
\begin{enumerate*}[label=(\arabic *)] 
\item Some studies also call the first-stage retrieval as a ranking stage, a search stage, or a recall stage. In this survey, we will refer to it as the retrieval stage for consistency and simplicity;
\item The survey mainly focuses on ranking algorithms of semantic retrieval models, thus will only briefly mention indexing methods. Readers who are interested in sparse or dense indexing techniques could refer to \cite{zobel2006inverted, muja2014scalable, chen2018sptag, zhang2019grip}.
\end{enumerate*}

So far as we know, this is the first survey on both traditional and neural semantic models for the first-stage retrieval. 
It reviews early semantic retrieval models proposed from 1990 to 2013, and covers neural semantic retrieval models published in major conferences (e.g.,  ACL, ICLR, AAAI, SIGIR, TheWebConf, CIKM, WSDM, EMNLP, and ECIR) and journals (e.g., TOIS, TKDE, TACL, and IP$\&$M) in the fields of deep learning, natural language processing and information retrieval from 2013 to June 2021.
There have been some surveys on neural models for IR~\cite{mitra2017neural, onal2018neural, mitra2018introduction, guo2019deep}, but none of them focused on the first-stage retrieval. For example, \citet{onal2018neural} paid attention to the application of neural methods to different IR tasks. \citet{guo2019deep} took a deep look into deep neural networks for re-ranking stages. For the first-stage retrieval, the booklet by \citet{li2014semantic} talked about early semantic retrieval models, but without recent booming neural models for the first-stage retrieval. Recently, \citet{lin2020pretrained} discussed several pre-training models for the first-stage retrieval and re-ranking stages. Different from them, we make an comprehensive overview of semantic models for the first-stage retrieval under a unified framework, including early semantic retrieval models, neural semantic retrieval models and the connection between them.

To sum up, our contributions include:
\begin{enumerate}
\item We describe the current landscape of the first-stage retrieval models under a unified framework to clarify the connection between the classical term-based retrieval, early methods for semantic retrieval and neural methods for semantic retrieval.
\item We provide a comprehensive and up-to-date review of semantic retrieval models, with a brief review of early semantic retrieval models and a detailed description of recent neural semantic retrieval models.
\item We summarize neural semantic retrieval models into three paradigms from the perspective of model architecture, i.e., sparse retrieval methods, dense retrieval methods and hybrid retrieval methods. We also discuss key topics on model learning, including loss functions and negative sampling strategies.
\item We discuss some open challenges and suggest potentially promising directions for future works.
\end{enumerate}

We organize this survey as follows. We first introduce three typical applications of semantic retrieval models in Section 2. Then, we provide some background knowledge, including problem formalization, index methods and classical term-based retrieval methods in Section 3. We sketch early methods for semantic retrieval in Section 4. In Section 5, we review existing neural methods for semantic retrieval from the perspective of model architecture, and introduce key topics on model learning. Finally, we discuss challenges and future directions in Section 6, and conclude this survey in Section 7.

\section{Major Applications of Semantic Retrieval Models}

The first-stage retrieval plays an essential role in almost all large-scale IR applications. In this section, we describe three major text retrieval applications, including ad-hoc retrieval~\cite{ricardo2011modern}, open-domain question answering~\cite{simmons1965answering, voorhees2000building}, and community-based question answering~\cite{burke1997question, srba2016a}.

Ad-hoc retrieval is a typical retrieval task, and there has been a long research history on ad-hoc retrieval models. In this task, users express their information needs as queries, then trigger searches in the retrieval system to obtain relevant documents. All retrieved documents are often returned as a ranked list according to the degree of relevance to the user query. A major characteristic of ad-hoc retrieval is the length heterogeneity between the query and the document. Queries are often short in length, consisting of only a few keywords~\cite{mitra2017neural}. While documents have longer texts, ranging from multiple sentences to several paragraphs. 
Such heterogeneity between queries and documents leads to the classical vocabulary mismatch problem, which has been a long-term challenge in both the retrieval stage as well as re-ranking stages in ad-hoc retrieval~\cite{li2014semantic}.
The earliest datasets to support reliable evaluation of the first-stage retrieval models are always based on TREC collections, such as Associated Press Newswire (AP), Wall Street Journal (WSJ) and Robust~\cite{kwok2004trec}. The number of documents in these collections is usually hundreds of thousands, and documents are usually news articles. Later, larger collections based on Web data, such as ClueWeb~\cite{clarke2009overview}, are built for the evaluation of retrieval technology. However, the number of queries in these datasets is only a few hundred, which is not enough for the training of neural-based retrieval models. In recent years, large-scale datasets, such as MS MARCO~\cite{nguyen2016msmarco}, TREC CAR~\cite{dietz2017trec} and TREC Deep Learning Track~\cite{craswell2020overview}, are released, which label relevant documents for hundreds of thousands of queries. The availability of these large-scale datasets has greatly promoted the development of neural retrieval models. 
Besides, there are also some domain-specific retrieval datasets, e.g., GOV2~\cite{clarke2004overview}, TREC Medical Records Track (MedTrack) and TREC-COVID~\cite{voorhees2021trec}, which are also commonly used for the evaluation.

Open-domain question answering (OpenQA) is a task to answer any sort of (factoid) questions that humans might ask, using a large collection of documents (e.g., Wikipedia, or Web page) as the information source \cite{karpukhin2020dense}. Unlike the ad-hoc retrieval which aims to return a ranked list of documents, the OpenQA task is to extract a text span as the answer to the question. To achieve this, most existing works build the OpenQA system as a two-stage pipeline \cite{chen2017reading}:
\begin{enumerate*}[label=(\arabic *)]
\item A \textit{document retriever} selects a small set of relevant documents that probably contain the answer from a large-scale collection;
\item A \textit{document reader} extracts the answer from relevant documents returned by the document retriever.
\end{enumerate*}
In our work, we only consider the document retriever component since the document reader is out of the scope of this paper. Typically, the question in OpenQA tasks is a natural language sentence, which has well-formatted linguistic structures. While the document is often a small snippet of text, ranging from several sentences to a passage \cite{dhingra2017quasar, dunn2017searchQA}. Moreover, relevant documents are required to be not only topically related to but also correctly address the question, which requires more semantics understanding except for exact term matching features. 
For the evaluation of the first-stage retrieval models on OpenQA tasks, several benchmark datasets are available. Most commonly used datasets, such as SQuAD-open~\cite{chen2017reading}, SearchQA~\cite{dunn2017searchQA}, TriviaQA-unfiltered~\cite{joshi2017triviaQA} and Natural Questions Open~\cite{kwiatkowski2019natural}, have tens of thousands of queries for model training. Several smaller-scale datasets, e.g., WebQuestions~\cite{berant2013semantic} and CuratedTREC~\cite{baudis2015modeling}, are also often used for model evaluation. The document collection in these datasets is usually based on Wikipedia pages (e.g., SQuAD-open and Natural Questions Open) or Web pages (e.g. SearchQA, and WebQuestions), and queries are written by crowd-workers (e.g., SQuAD-open) or crawled from existing websites (e.g., SearchQA and TriviaQA-unfiltered).

Community-based question answering (CQA) aims to address user's questions using the archived question-answer (QA) pairs in the repository, since CQA systems have already accumulated a large amount of high-quality human-generated QA pairs, such as Yahoo! Answers\footnote{\url{https://answers.yahoo.com}}, Stack Overflow\footnote{\url{http://www.stackoverflow.com/}} and Quora\footnote{\url{http://www.quora.com/}}. There are two different ways to produce the answer to a user's question. One is to directly retrieve answers from the collection if the answer exists \cite{wan2016match}. The other is to select the duplicate question from the collection and take the accompanied answer as the result~\cite{wang2020match2}. 
Both of these two ways require the retrieval system to firstly recall a subset of candidates from the whole collection, and then re-rank candidates to generate the final result. However, targets (i.e., answers and questions) in these two ways often have very different expressions, leading to different challenges in terms of semantic modeling. Firstly, the duplicate question retrieval needs to capture semantic similarities between words (phrases) since there are often different ways to express the same question. Secondly, the answer retrieval needs to model logical relationships between questions and answers. 
Although many datasets are constructed based on CQA data, few of them are suitable for evaluating the first-stage retrieval models. Existing related works usually conduct experiments on QQP\footnote{\url{https://data.quora.com/First-Quora-Dataset-ReleaseQuestion-Pairs}} and WikiAnswers~\cite{fader2013paraphrase} datasets.

There are also some other retrieval scenarios, such as entity linking~\cite{gillick2019learning}, e-commerce search~\cite{li2019semantic, zhang2020towards, li2019multi} and sponsored search~\cite{fan2019mobius}. For these applications, academic researchers and industrial developers have realized the importance of utilizing semantic information for the first-stage retrieval. Due to page limitations, we will not discuss these works in this survey, but it is possible and necessary to generalize techniques applied in text retrieval to other retrieval tasks.

\section{Background}
In this section, we first characterize the first-stage retrieval by giving a unified formulation of the first-stage retrieval models. Then, we introduce typical indexing methods cooperating retrieval models to support efficient retrieval. Finally, we summarize classical term-based retrieval methods.

\subsection{Problem Formalization}
Given a query $q$, the first-stage retrieval aims to recall all potentially relevant documents from a large corpus $\mathcal{C} = \left\{d_1, d_2, \cdots, d_N\right\}$. Different from re-ranking stages with a small set of candidates, the corpus size $N$ for the first-stage retrieval can range from millions (e.g., Wikipedia) to billions (e.g., the Web). Thus, efficiency is a crucial concern for models used in the first-stage retrieval.

Formally, given a dataset $\mathcal{D}=\left\{\left(q_i, D_i, Y_i\right)\right\}_{i=1}^{n}$, where $q_i$ denotes a user query, $D_i {=} [d_{i1}, d_{i2},\cdots, d_{ik}]$ denotes a list of documents to the query $q_i$, and $Y_i = [{y_{i1}, y_{i2}, \cdots, y_{ik}}] \in \{1, 2, \cdots, l\}$ is the corresponding relevance label of each document in $D_i$. There exists a total order between relevance labels $l \textgreater l-1 \textgreater \cdots \textgreater 1$, where $\textgreater$ denotes the order relation. Note here the number of labeled documents $k$ to each query is often significantly smaller than the corpus size $N$, since it is impossible to manually annotate all the huge amount of documents. The goal of the first-stage retrieval is to learn a model $s(\cdot, \cdot)$ from $\mathcal{D}$ that gives high scores to relevant $(q, d)$ pairs and low scores to irrelevant ones. For any query-document pair $(q, d)$, $s(q, d)$ gives a score that reflects the relevance degree between $q$ and $d$, and thus allows one to rank all the documents in the corpus $\mathcal{C}$ according to predicted scores. 
Without loss of generality, the scoring function can be abstracted by the following unified formulation:
\begin{equation}
s(q, d) = f\bigl(\phi(q), \psi(d)\bigr),  \label{unified}
\end{equation}
where $q \in X$ and $d \in Y$ are the input query and document, and two representation functions $\phi: X \rightarrow \mathbb{R}^{k_1}$ and $\psi: Y \rightarrow \mathbb{R}^{k_2}$ map a sequence of tokens in $X$ and $Y$ to their associated embeddings $\phi(q)$ and $\psi(d)$, respectively.
To build a responsive model for the first-stage retrieval, it leads to a number of requirements on these three components:
\begin{itemize}
\item The document representation function $\psi$ should be independent of the query since queries are unknown before the search system is deployed. In this way, document representations can be pre-computed and indexed offline with methods in Section \ref{index}. Meanwhile, this means that the $\psi(d)$ component can be sophisticated to some extent since it has no impact on the online serving.
\item The query representation function $\phi$ is required to be as efficient as possible since it needs to compute query embeddings online. Thanks to the nature of independence, two components $\phi$ and $\psi$ can be identical or different, which is flexible enough to design models for different retrieval tasks with homogeneous or heterogeneous inputs.
\item  To satisfy the real-time retrieval requirement, on the one hand the scoring function $f$ should be as simple as possible to minimize the amount of online computation, and on the other hand, it must take the indexing method into account.
\end{itemize}

\subsection{Indexing Methods} \label{index}
As mentioned above, one major difference between the first-stage retrieval and re-ranking stages is that the former does ranking on large-scale documents in the repository. Thus, the efficiency of the first-stage retrieval models is one of the core considerations. In practice, to support storing and fast retrieval of documents in the whole repository, retrieval systems need to build an index, where the indexing technique is crucial to the rapid response during the online serving. There are many indexing techniques, such as signature, inverted index, and dense vector index. Rather than exploring all the existing approaches, we only describe the fundamental principle of two typical indexing schemes.

The inverted index is currently the most popular indexing scheme and is used for many applications due to its simplicity and efficiency. Before building an inverted index, each document in the collection is parsed and segmented into a list of tokens. Then, the inverted index is created, which mainly consists of a dictionary and a collection of posting lists. The dictionary includes all the terms found in the collection and their document frequencies. Each posting list records document identifiers, term occurrence frequencies, and possibly other information of documents in which the corresponding term appears. During the online serving, for a user's query, top $k$ most similar documents are fetched in turn with the help of the inverted index. Concretely, the query is processed with one term at a time. Initially, each document has a similarity of zero to the query. Then, for each query term $t$, the similarity score of each document in $t$’s posting list increases by the contribution of $t$ to the similarity of the query-document pair. Once all query terms have been processed, the $k$ largest similarity scores are identified, and the corresponding document list is returned to the user. In fact, many acceleration strategies are applied during the process to improve retrieval efficiency, but they are omitted here. More details about the inverted index technique could be found in~\cite{witten1999managing, zobel2006inverted}.

Along with the development of neural representation learning methods, dense vector index based on approximate nearest neighbor search algorithms is used to support the new representation paradigm. 
One of reasons why the inverted index works well is that the documents–term matrix is very sparse. However, most semantic retrieval models produce dense and distributed document representations, thus the inverted index method is no longer feasible to retrieve documents efficiently from a large collection.
From the equation \eqref{unified}, the retrieval problem could be viewed as the nearest neighbor search problem \cite{shakhnarovich2006nearest}, once the query embedding and all the document embeddings have been calculated. This fundamental problem has been well studied in the research community \cite{abbasifard2014survey, andoni2009nearest}. The simplest approach to the nearest neighbor search is the brute-force search, which scans all the candidates and computes similarity scores one by one. However, the brute-force search becomes impractical when the size of collections exceeds a certain point. Thus, most researches resort to an approximate nearest neighbor (ANN) search~\cite{aumuller2017ann, echihabi2020return, li2019approximate}, which allows for a slight loss in precision while yielding multiple orders of magnitude improvement in speed. Generally, existing ANN search algorithms can be categorized into four major types, including tree-based~\cite{bentley1975multidimensional, beis1997shape}, hashing-based~\cite{indyk1998approximate, datar2004locality}, quantization-based~\cite{jegou2010product, ge2013optimized} and proximity graph approaches~\cite{kleinberg2000navigation, malkov2018efficient}. The earliest solutions to ANN search are based on locality-sensitive hashing~\cite{indyk1998approximate}, but currently proximity graph methods~\cite{kleinberg2000navigation, malkov2018efficient} yield a better performance among all the approaches in most respects based on a popular benchmark\footnote{\url{http://ann-benchmarks.com/}}. Graph-based methods build the index by retaining the neighborhood information for each individual data point towards other data points or a set of pivot points. Then, various greedy heuristics are proposed to navigate the proximity graph for a given
query point. So far, several open-source libraries for ANN search, such as Faiss~\cite{JDH17} and SPTAG~\cite{ChenW18}, have been developed, and search engines\footnote{\url{https://www.elastic.co/cn/elasticsearch/}}\textsuperscript{,}\footnote{\url{https://vespa.ai/}}\textsuperscript{,}\footnote{\url{https://milvus.io/}} supporting ANN search have been built and applied widely.

\subsection{Classical Term-based Retrieval}
This subsection provides an overview of classical term-based methods for the first-stage retrieval, including the vector space model, probabilistic retrieval models and language models for IR. 
In general, these methods build representations of queries and documents based on the bag-of-words (BoW) assumption where each text is represented as a bag (multiset) of its words, disregarding grammar and even word order.
Particularly, the representation functions $\phi$ and $\psi$ are set to be manually defined feature functions, such as word frequency, and dimensions of representations (i.e., $k_1$ and $k_2$) are generally equal to the vocabulary size. The representation functions $\phi$ and $\psi$ are usually different for queries and documents, but they all pledge the sparsity of representations so that the inverted index could be used to support efficient retrieval.

The early representative of term-based methods is the vector space model (VSM)~\cite{salton1975vector} which represents queries and documents as high-dimensional sparse vectors in a common vector space. Under this framework, queries and documents are viewed as vectors with each dimension corresponding to a term in the vocabulary, where the weight of each dimension can be determined by different functions, e.g., term frequency (TF), inverse document frequency (IDF) or the composite of them~\cite{salton1991developments, salton1988term}. Then, one can use the similarity (usually cosine similarity) between a query vector and a document vector as the relevance measure of the query-document pair. The resulting scores can then be used to select the top relevant documents for the query. 
The VSM has become the fundamental of a series of IR solutions---the probabilistic retrieval model and language model for IR can be both viewed as the instantiation of VSM with different weighting schemes.

Probabilistic methods are one of the oldest formal models in IR, which introduce the probability theory as a principled foundation to estimate the relevance probability $P(y=1|q,d)$ of a document $d$ to the query $q$. 
The Binary Independence Model (BIM)~\cite{robertson1976relevance} is the most original and influential probabilistic retrieval model. 
It represents documents and queries to binary term vectors, that an entry is 1 if the corresponding term occurs in the document, and otherwise the entry is 0. With these representations, ``binary'' and ``term independency'' assumptions are introduced by BIM. But these assumptions are contrary to facts, so a number of extensions are proposed to relax some assumptions of BIM, such as the Tree Dependence Model~\cite{van1977theoretical} and BM25~\cite{Robertson:Now:BM25}. In particular, the BM25 model takes into account the document frequency, document length and term frequency, which has been widely used and quite successful across different academic researches as well as commercial systems~\cite{pedersen2010query, liu2017cascade}.

Instead of modeling the relevance probability explicitly, language models (LM) for IR~\cite{ponte1998language} build a language model $M_d$ for each document $d$, then documents are ranked based on the likelihood of generating the query $q$, i.e.,  $P(q|M_d)$. The document language model is also built on the bag-of-words assumption, and could be instantiated as either multiple Bernoulli~\cite{ponte1998language} or Multinomial \cite{Miller1999A, hiemstra2000probabilistic}. 
Experimental results in~\cite{ponte1998language} prove the effectiveness of term weights coming from language models over the traditional TF-IDF weight. Moreover, language models provide another perspective for modeling retrieval tasks, and subsequently inspire many extended approaches~\cite{zhai2008statistical, bravo2010hypergeometric}. 

In summary, modeling relevance in a shallow lexical way, especially combined with the inverted index, endows classical term-based models a key advantage on efficiency, making it possible to retrieve from billions of documents quickly. However, such a paradigm is also accompanied by clear drawbacks, like the well-known vocabulary mismatch problem or not well capturing text semantics.
Therefore, more sophisticated semantic models for improving the first-stage retrieval performance start to attract researchers' interests in the following.

\section{Early methods for semantic retrieval}
From the 1990s to the 2000s, extensive studies have been carried out to improve term-based retrieval methods. 
Most of them mine information from external resources or the collection itself to enrich query representations $\phi(q)$, document representations $\psi(d)$ or both of them for semantic retrieval.
Here, we sketch a brief picture of some of them.

\subsection{Query Expansion}
To compensate for the mismatch between queries and documents, the query expansion technique is used to expand the original query with terms selected from external resources~\cite{xu2017quary}.
In this way, query representations $\phi(q)$ are enriched, and more documents could be considered during the retrieval process through the extended query terms.

Query expansion is the process of adding relevant terms to a query to improve retrieval effectiveness. There are a number of query expansion methods, and they can be classified into global methods~\cite{lesk1969word, qiu1993concept} and local methods~\cite{abdul2004umass, zhai2001model}.
Global methods expand or reformulate query words by analyzing word co-occurrences from the corpus being searched or using an external hand-crafted thesaurus (e.g., WordNet)~\cite{Voorhees1994Query}. Although a number of data-driven query expansion methods, such as~\cite{Bai2007Using}, can improve the average retrieval performance, they are shown to be unstable across queries. On the other hand, local methods adjust a query based on top-ranked documents retrieved by the original query. 
This kind of query expansion is called pseudo-relevance feedback (PRF)~\cite{cao2008selecting}, which has been proven to be highly effective to improve the performance of many retrieval models~\cite{lv2009comparative, rocchio1971relevance}. Relevance model~\cite{lavrenko2017relevance}, mixture model, and divergence minimization model~\cite{zhai2001model} are the first PRF methods proposed under the language modeling framework. Since then, several other local methods have been proposed, but the relevance model is shown to be still among state-of-the-art PRF methods and performs more robustly than many other methods~\cite{lv2009comparative}.

In general, query expansion methods have been widely studied and adopted in IR applications, but they do not always yield a consistent improvement. Especially expansion methods based on pseudo-relevance feedback are prone to the query drift problem~\cite{collins2009reducing}.
Subsequently, with the development of deep learning technique, neural word embeddings and deep language models are used to enhance query expansion methods~\cite{roy2016using, diaz2016query, mao2020generation}.

\subsection{Document Expansion} \label{Document Expansion}
An alternative to query expansion is to perform the expansion for all documents in the corpus, then those enriched documents are indexed and searched as before. Intuitively, document expansion methods supplement each posting list in the inverted index, which have shown to be particularly effective for information retrieval tasks~\cite{efron2012improving, agirre2010document, tao2006language}.

Document expansion is first proposed in the speech retrieval community~\cite{singhal1999document}. \citet{singhal1999document} proposed to use the original document as a query into the collection, and the ten most relevant documents were selected. Then, they enhanced the representation of the original document by adding to the document vector a linearly weighted mixture of related documents. Similarly, \citet{efron2012improving} followed a similar approach on short text retrieval tasks. They submitted documents as pseudo-queries and performed document expansion based on the analysis of the result set.
Different from the retrieval-based method to determine related documents for expansion, it is another way to use document clustering to determine similar documents, and document expansion is carried out with respect to these results~\cite{liu2004cluster, kurland2004corpus}. Both works report significant improvements over non-expanded baselines on the TREC ad-hoc document retrieval task. 
In addition to using the document collection itself, it is also helpful to use external information to augment document representations~\cite{sherman2017document, agirre2010document}. 
For example, \citet{agirre2010document} presented a novel document expansion method based on a WordNet-based system to find related concepts and words, which is the first to perform document expansion using lexical semantic resources.

Document expansion technique has been less popular with IR research because they are less amenable to rapid experiments. The corpus needs to be re-indexed every time the expansion technique changes, which is a costly process. In contrast, manipulations to query representations can happen at retrieval time and hence are much faster. Besides, the success of document expansion has been mixed. \citet{billerbeck2005document} explored both query expansion and document expansion in the same framework and concluded that the former is consistently more effective.
Nevertheless, dramatic improvement for the first-stage retrieval has been achieved after equipping the document expansion technique with neural models, such as doc2query~\cite{nogueira2019document} and docTTTTTquery~\cite{nogueira2019doc2query} (See Section \ref{sparse_retrieval_methods}).

\subsection{Term Dependency Models}
Typically, term-based methods consider terms in the document independently and ignore the term orders.
As a result, concepts represented by multiple contiguous words cannot be depicted correctly, and the stronger relevance of consecutive or ordered terms matching between queries and documents cannot be reflected well.
Term dependency models attempt to address the above problem by incorporating term dependencies into the representation functions $\phi$ and $\psi$.

A natural way is to extend the dictionary in the inverted index with frequent phrases. For example, \citet{fagan1987experiments} tried to incorporate phrases into the VSM, where phrases are viewed as additional dimensions in the representation space. Then, the scoring function can be formalized to the combination of term-level score and phrase-level score:
\begin{equation}
s(q, d) = w_{\text{term}} \cdot \underbrace{s_{\text{term}}(q, d)}_{\text{term score}} +\, w_{\text{phr}} \cdot \underbrace{s_{\text{phr}}(q, d)}_{\text{phrase score}},
\end{equation}
where $w_{\text{term}}$ and $w_{\text{phr}}$ are weights to achieve weight normalization, and the score of a phrase can be defined as the average of TF-IDF weights of its component terms.
\citet{xu2010relevance} also investigated the approach that extends BM25 with n-grams. They defined the BM25 kernel as follows:
\begin{equation}
\operatorname{BM25-Kernel}(q,d)=\sum_{t} \operatorname{BM25-Kernel}_{t}(q,d),
\end{equation}
where $\operatorname{BM25-Kernel}_{t}(q, d)$ denotes the BM25 kernel of type $t$, and $t$ can be bigram, trigram, etc.
\begin{equation}
\operatorname{BM25-Kernel}_t(q, d)= \sum_x  \operatorname{IDF}_t(x) \times \frac{(k_3 + 1) \times f_t(x, q)}{k_3+f_t(x, q)}  \times \frac{(k_1 + 1) \times f_t(x, d)}{k_1\left(1 - b + b \frac{f_t(d)}{\bar{f_t}}\right) + f_t(x, d)},
\end{equation}
where $x$ denotes a n-gram of type $t$, $f_t(x, q)$ and $f_t(x, d)$ are frequency of unit $x$ in query $q$ and document $d$ respectively, $f_t(d)$ is total number of units with type $t$ in document $d$, $\bar{f_t}$ is average number of $f_t(d)$ within the whole collection, $k1$, $k3$, and $b$ are parameters.

Integrating term dependencies to term-based methods increases the complexity, but gains are not significant as expected~\cite{lavrenko2008generative}. The Markov Random Field (MRF) approach proposed by \citet{metzler2005markov} reports the first clear improvement for term dependency models over term-based baselines. In MRF, the document and each term in the query are represented as a node respectively. The document node is connected to every query term node. Moreover, there are some edges between query term nodes, based on pre-defined dependency relations (e.g., bigram, named entity, or co-occurrence within a distance), to represent their dependencies. Then, the joint probability of query $q$ and document $d$ can be formally represented as
\begin{equation}
P(q, d)=\frac{1}{Z} \prod_{c \in \operatorname{clique}(\mathrm{G})} \exp \bigl(\lambda_{c} f(c)\bigr),
\end{equation}
where $c$ denotes a clique on the constructed graph $\mathrm{G}$, $\lambda_c$ is the interpolation coefficient, $f(c)$ is the potential function defined on clique $c$, and $Z$ denotes the partition function. 
In practice, we can define different feature functions to capture different types of term dependencies, and the coefficient $\lambda_c$ can be optimized towards designative retrieval metrics, as in~\cite{metzler2005markov}.

While those methods are capable of capturing certain syntactics and semantics, their ``understanding'' capability is much limited.
How to go beyond these simple counting statistics and mine deeper signals to better query-document matching is still an open question.
Nevertheless, there is no doubt that term dependency models demonstrate the importance of understanding document semantics with context, stimulating a series of neural retrieval models that emphasize the capturing of contextual information~\cite{zhan2020repBERT, khattab2020colbert}.

\subsection{Topic Models}  \label{Topic Models}
Another line to improve $\phi$ and  $\psi$ simultaneously focuses on semantic relationships between words---usually modeling words' co-occurrence relation to discover latent topics in texts and matching queries and documents by their topic representations. 
In this way, each dimension of the representation indicates a topic instead of a term. Besides, the inverted index becomes impractical since topic representations lose sparsity.

Topic modeling methods have received much attention in natural language processing (NLP) tasks. Overall, they can be divided into two categories, including probabilistic and non-probabilistic approaches. 
The non-probabilistic topic model, such as latent semantic indexing (LSI)~\cite{deerwester1990indexing}, non-negative matrix factorization (NMF)~\cite{lee2001algorithms}, and regularized latent semantic indexing (RLSI)~\cite{wang2011regularized}, is usually obtained by matrix factorization. Taking LSI as an example, it uses a truncated singular value decomposition (SVD) to obtain a low-rank approximation to the document-term matrix, then each document can be represented as a mixture of topics. 
Other topic models choose different strategies to conduct the matrix factorization. 
For example, NMF introduces the non-negative constraint and RLSI assumes topics are sparse.
For probabilistic approaches, probabilistic latent semantic indexing (pLSI)~\cite{hofmann1999probabilistic} and latent dirichlet allocation (LDA)~\cite{blei2003latent} are most widely used. Probabilistic topic models are usually generative models, where each topic is defined as a probabilistic distribution over terms in the vocabulary and each document in the collection is defined as a probabilistic distribution over topics.

Studies that apply topic models to improve retrieval results can be classified in two ways. The first is to obtain query and document representations in the topic space, and then calculate relevance scores based on topic representations. For example, the LSI learns a linear projection that casts the sparse bag-of-words text vector into a dense vector in latent topic space, then the relevance score between a query and a document is the cosine similarity of their corresponding dense vectors. In the LDA-based retrieval model~\cite{wei2006lda}, queries and documents are represented by their latent topic distributions. The relevance score of each query-document pair is computed by the Kullback-Leibler divergence as follows: 
\begin{equation}
s(q, d) = 1 - \frac{1}{2}\Bigl(\mathrm{KL}\left(v_q \| v_d\right) + \mathrm{KL}\left(v_d \| v_q\right)\Bigr) = 1 -\frac{1}{2} \sum_{k=1}^{K}\left(\left(v_q^k - v_d^k\right) \log \frac{v_q^k}{v_d^k}\right),
\end{equation}
where $v_q$ and $v_d$ are topic representations of query $q$ and document $d$ respectively, and $v_q^k$ and $v_d^k$ are the $k$-th element of $v_q$ and $v_d$.

Another way is to combine topic models with term-based methods. A simple and direct approach is to linearly combine relevance scores calculated by topic models and term-based models~\cite{hofmann1999probabilistic}:
\begin{equation}
s(q, d) = \alpha s_{\text{topic}}(q, d) + (1-\alpha) s_{\text{term}}(q, d),
\end{equation}
where $\alpha$ is the coefficient, $s_{\text{topic}}(q, d)$ and $s_{\text{term}}(q, d)$ are the topic matching score and term matching score respectively. 
In addition, probabilistic topic models can be taken as the smoothing method to language models for IR~\cite{diaz2005regularizing, wei2006lda, yi2009comparative}.
\begin{equation}
P(q|d) = \prod_{w \in q} P(w|d) = \prod_{w \in q}\Bigl(\alpha P_{\operatorname{LM}}(w|d) + (1-\alpha) P_{\operatorname{TM}}(w|d)\Bigr),
\end{equation}
where $\alpha$ is the coefficient, $P_{\operatorname{LM}}(w|d)$ and $P_{\operatorname{TM}}(w|d)$ are generating probabilities of word $w$ given document $d$ estimated by a language model and a topic model. 
The $P_{\operatorname{TM}}(w|d)$ can be defined as:
\begin{equation}
P_{\operatorname{TM}}(w|d) = \sum_{z=1}^K P(w|z) P(z|d),
\end{equation}
where $z$ denotes a latent topic.

According to results in~\cite{atreya2011latent}, using latent topic representations obtained by topic models alone for IR tasks only has small gains or poor performance over term-based baselines, unless combining them with term-based methods. Possible reasons include:
\begin{enumerate*}[label=(\arabic *)]
\item These topic models are mostly unsupervised, learning with a reconstruction objective, either based on mean squared error~\cite{deerwester1990indexing} or likelihood~\cite{hofmann1999probabilistic, blei2003latent}. They may not learn a matching score that works well for specific retrieval tasks; 
\item Word co-occurrence patterns learned by these topic models are from documents, ignoring the fact that language usages in searching texts (queries) can be different from those in writing texts (documents), especially when the heterogeneity between queries and documents is significant;
\item Topic models represent documents as compact vectors, losing detailed matching signals over term-level. 
\end{enumerate*}
Later, using more powerful neural models, e.g., doc2vec~\cite{le2014distributed}, instead of topic models for information retrieval has achieved better results~\cite{ai2016improving, ai2016analysis}.

\subsection{Translation Models}
A notable attempt to address the vocabulary mismatch problem is the statistical translation approach, which enriches the document representation function $\psi$ from term frequency to translation models.
Statistical machine translation (SMT) is leveraged for IR by viewing queries as texts in one language and documents as texts in another language. Retrieval by translation models needs to learn translation probabilities from queries to associated relevant documents, which can be obtained from labeled data, and thus belongs to the supervised learning approach.

\citet{berger1999information} firstly proposed to formulate retrieval tasks as SMT problem, in which query $q$ is translated into document $d$ with the conditional probability $P(d|q)$. The model can be written as:
\begin{equation}
    P(d|q) \propto P(q|d) P(d),
\end{equation}
where $P(q|d)$ denotes a translation model which translates $d$ to $q$, and $P(d)$ denotes a language model giving rise to $d$. Translation probabilities can be estimated with queries and their associated relevant documents, e.g., click-through datasets, and the language model can be learned with different schemes, such as BM25. As \citeauthor{karimzadehgan2010estimation} have noted~\cite{karimzadehgan2010estimation}, the translation probability $P(q|d)$ allows for the incorporation of semantic relations between terms with non-zero probabilities, which provides a sort of ``semantic smoothing'' for $P(q|d)$.

One important difference between conventional machine translation and machine translation for retrieval is that both queries (target language) and documents (source language) are in the same language. The probability of translating a word to itself should be quite high, i.e., $P(w \mid w) > 0$, which corresponds to exact term matching in retrieval tasks. How to accurately calculate self-translation probabilities is an important issue. If self-translation probabilities are too large, it will make other translation probabilities small and decrease the effect of using translation. On the other hand, if self-translation probabilities are too small, then it will make exact matching less effective and hurt the performance of retrieval. A number of methods~\cite{gao2010clickthrough, karimzadehgan2010estimation, karimzadehgan2012axiomatic} have been proposed to estimate self-translation probabilities. For example, \citet{karimzadehgan2010estimation} proposed to address this estimation problem based on normalized mutual information between words, which is less computationally expensive and has better coverage of query words than the synthetic query method of estimation~\cite{berger1999information}:
\begin{equation}
P_{\mathit{mi}-\alpha}=\left\{\begin{array}{ll}\alpha+(1-\alpha) P_{\mathit{mi}}(w \mid u) & \text { if } w=u \\ (1-\alpha) P_{\mathit{mi}}(w \mid u) & \text { if } w \neq u\end{array}\right.
\end{equation}
where $\alpha$ is the weight which is empirically set on heldout data.
Similarly, an alternative heuristic is to impose constant self-translation probabilities for all words in the vocabulary~\cite{karimzadehgan2012axiomatic}, i.e., setting $P(u|u)$ to a constant value $s$ for every $u$, where $P_t(w|u)$ is estimated according to:
\begin{equation}
P_{\mathit{mi}-s}=\left\{\begin{array}{ll}s & \text { if } w=u \\ (1-s) \frac{P_{\mathit{mi}}(w \mid u)}{\sum_{v \neq u} P_{\mathit{mi}}(v \mid u)} & \text { if } w \neq u\end{array}\right.
\end{equation}
All these methods assume that self-translation probabilities estimated directly from data are not optimal for retrieval tasks, and the authors have demonstrated that significant improvement can be achieved by adjusting the probabilities~\cite{gao2010clickthrough}. 

Statistical translation models have also been applied to query expansion. For example, \citet{riezler2010query} suggested utilizing a word-based translation model for query expansion. The model is trained with click-through data consisting of queries and snippets of clicked web pages. \citet{gao2012towards} generalized the word-based translation model to a concept-based model and employed the model in query expansion.

Nevertheless, SMT models have not been used much because they are difficult to train due to data sparsity, and are not more effective than the term-based retrieval with pseudo-relevance feedback~\cite{lavrenko2017relevance} in most situations. Subsequently, after the appearance of neural word embeddings, using distributed representations to calculate translation probabilities and improve translation models are proposed naturally~\cite{ganguly2015word, zuccon2015integrating}.

\vbox{}
\textbf{Takeaway.}
Early semantic retrieval models, such as query expansion, document expansion, term dependency models, topic models, and translation models, aim to improve classical BoW representations with semantic units extracted from external resources or the collection itself. Most of them still follow classical term-based methods by representing texts with high-dimensional sparse vectors in symbolic space, so as to be easily integrated with the inverted index to support efficient retrieval. However, these approaches always rely on hand-crafted features to build representation functions. As a result, only shallow syntactic and semantic information can be captured.
Nevertheless, these early proposals are crucial because they have initially explored beneficial factors for the first-stage retrieval. Thereby, a series of new semantic retrieval models could be inspired when the deep learning technique breaks out, and exciting results could be obtained concomitantly.

\section{Neural methods for semantic retrieval}
\label{Neural methods for semantic retrieval}

\begin{table}
\caption{Overview of Neural Methods for Semantic Retrieval.}
\label{tab:summary}
\renewcommand{\arraystretch}{1.0}
\centering
\begin{adjustbox}{max width=\textwidth}
\begin{tabular}{l l l c c c}
\toprule
\multirow{2}{*}{ } &\multicolumn{2}{c}{ Model } &\multicolumn{3}{c}{Task} \\
\cmidrule(lr){2-3} \cmidrule(lr){4-6}
& Type & Representative Work & Ad-hoc Retrieval & OpenQA & CQA \\ 
\midrule
\multirow{17}{*}{\makecell[l]{Sparse Retrieval Methods}} 
&\multirow{13}*{\makecell[l]{Neural Weighting Schemes} }
& DeepTR~\cite{zheng2015learning} &$\surd$ &  &  \\ 
& &NTLM~\cite{zuccon2015integrating} &$\surd$ &  &  \\ 
& &TVD~\cite{frej2020learning} &$\surd$ & & \\
& &DeepCT~\cite{dai2019context, dai2020context} & $\surd$ & & \\
& &HDCT~\cite{dai2020context_www} &$\surd$  & & \\
& &\citet{mitra2019incorporating} &$\surd$  &  & \\ 
& &\citet{mitra2020conformer} &$\surd$ & & \\
& &GAR~\cite{mao2020generation} & & $\surd$ & \\
& &doc2query~\cite{nogueira2019document} & $\surd$ & & \\
& &docTTTTTquery~\cite{nogueira2019doc2query} & $\surd$ & & \\
& &UED~\cite{yan2021unified} & $\surd$ & & \\
& &SparTerm~\cite{bai2020sparterm} & $\surd$  & & \\
& &DeepImpact~\cite{mallia2021learning} & $\surd$  & & \\
\cmidrule{2-6}
&\multirow{4}*{\makecell[l]{Sparse Representation Learning} }
& Semantic Hashing~\cite{salakhutdinov2009semantic} &$\surd$& & \\
& &SNRM~\cite{zamani2018neural} &$\surd$ & & \\
& &UHD-BERT~\cite{jang2021uhd}  & $\surd$ & & \\
& &\citet{ji2019efficient} & $\surd$ &  & \\
\midrule
\multirow{31}{*}{\makecell[l]{Dense Retrieval Methods}} 
&\multirow{9}*{\makecell[l]{Term-level Representation Learning} }
& OoB~\cite{Kenter2015Short} &$\surd$  &  &  \\ 
& &DESM~\cite{mitra2016dual} &$\surd$  &  &  \\ 
& &DC-BERT~\cite{zhang2020dc} &  &$\surd$  &  \\ 
& &ColBERT~\cite{khattab2020colbert} & $\surd$ & & \\
& &COIL~\cite{gao2021coil} & $\surd$ & & \\
& &De-Former~\cite{cao2020deformer} &  &$\surd$  & \\ 
& &PreTTR~\cite{macavaney2020efficient} &$\surd$  &  & \\ 
& &PIQA~\cite{seo2018phrase} & & $\surd$ & \\
& &DenSPI~\cite{seo2019real} &  &$\surd$ &  \\
& &SPARC~\cite{lee2020contextualized} & &$\surd$ &  \\
& &MUPPET~\cite{feldman2019multi} & &$\surd$ & \\
\cmidrule{2-6}
&\multirow{22}*{\makecell[l]{Document-level Representation Learning}}
& FV~\cite{clinchant2013aggregating} &$\surd$ & & \\
& &\citet{gillick2018end} &  &  &$\surd$ \\
& &\citet{ai2016analysis} & $\surd$  &  & \\
& &NVSM\cite{gysel2018neural} & $\surd$  &  &  \\
& &SAFIR~\cite{agosti2020learning} & $\surd$  &  &  \\
& &\citet{liu2016constraining} & $\surd$  &  & \\
& &\citet{tamine2019offline} & $\surd$  &  &  \\
& &\citet{henderson2017efficient} & &$\surd$ & \\
& &DPR~\cite{karpukhin2020dense} & &$\surd$ & \\
& &RepBERT~\cite{zhan2020repBERT} &$\surd$ & & \\
& &\citet{lin2020distilling} &$\surd$ & & \\
& &\citet{tahami2020distilling} & & & \\
& &DSSM~\cite{huang2013learning} &$\surd$ & & \\
& &ARC-I~\cite{hu2014convolutional} &$\surd$ & & \\
& &QA\_LSTM~\cite{tan2015lstm} &$\surd$ & & \\
& &ORQA~\cite{lee2019latent} & &$\surd$ & \\
& &REALM~\cite{guu2020realm} & &$\surd$ & \\
& &\citet{chang2020pre} & &$\surd$ & \\
& &\citet{liang2020embedding} &$\surd$ &$\surd$ & \\
& &Poly-encoders~\cite{humeau2019poly} &$\surd$ &$\surd$ & \\
& &ME-BERT~\cite{luan2020sparse} &$\surd$ &$\surd$ & \\
& &\citet{tang2021improving} & $\surd$ & $\surd$ & \\
\midrule
\multirow{11}{*}{\makecell[l]{Hybrid Retrieval Methods}}
& &\citet{vulic2015monolingual} & $\surd$  &  & \\
& &GLM~\cite{ganguly2015word} & $\surd$ &  & \\
& &DESM$_{MIXTURE}$~\cite{mitra2016dual} & $\surd$ &  & \\
& &\citet{roy2016representing} & $\surd$  &  &  \\
& &BOW-CNN~\cite{dos-santos-etal-2015-learning} & & & $\surd$  \\ 
& &EPIC~\cite{MacAvaney2020Expansion} & $\surd$ & & \\
& &DenSPI~\cite{seo2019real} &  &$\surd$ &  \\
& &SPARC~\cite{lee2020contextualized} & &$\surd$ &  \\
& &Hybrid~\cite{luan2020sparse}  & $\surd$  & $\surd$ &  \\
& &CLEAR~\cite{gao2020complementing} & $\surd$  &  &  \\
& &\citet{kuzi2020leveraging} &$\surd$  &  &  \\
\bottomrule
\end{tabular}
\end{adjustbox}
\end{table}

During the past decade, big data and fast computer processors have brought a new era for deep learning technique. A set of simple math units, called neurons, are organized into layers, and stacked into neural networks. Neural networks have the expressive power to represent complex functions and fit hidden correlations in complicated tasks~\cite{hundi2019deep}. For example, it converts discrete symbols (e.g., words, phrases and sentences) into low-dimensional dense vectors which are able to capture semantic and syntactic features for various NLP tasks~\cite{cui-etal-2017-attention, yao2019graph}. Naturally, it also attracts researchers from the IR field and leads to the research wave of neural approaches to IR (neural IR). However, most earlier researches focus on re-ranking stages~\cite{huang2013learning, guo2016deep}. Until recently, much attention is paid to explore neural networks to improve the semantic matching for the first-stage retrieval. 

Different from early semantic retrieval models, neural semantic retrieval models employ neural networks to build the representation functions (i.e., $\phi$ and/or $\psi$) as well as the scoring function (i.e., $f$). In this way, these models can learn deep semantics and complex interactions from data in an end-to-end way.
From the perspective of model architecture, neural methods for semantic retrieval can be categorized into three classes, including \textit{sparse retrieval methods}, \textit{dense retrieval methods}, and \textit{hybrid retrieval methods}. In this section, we will review major works about them. Table \ref{tab:summary} summaries surveyed neural semantic retrieval models in different categories.

\subsection{Sparse Retrieval Methods}\label{sparse_retrieval_methods}
Sparse retrieval methods usually represent each document and each query with sparse vectors, where only a small number of dimensions are active. 
The sparse representation has attracted great attention as it connects to the nature of human memories and shows better interpretability~\cite{bai2020sparterm}.
Besides, sparse representations can be easily integrated into existing inverted indexing engines for efficient retrieval.
Without loss of generality, sparse retrieval methods can be categorized into two classes. 
One is to encode queries and documents still in the symbolic space but employ neural models to improve term weighting schemes, namely \textit{neural weighting schemes}. The other is to directly learn sparse representations, i.e., $\phi(q)$ and $\psi(d)$, in latent space for queries and documents with neural networks, which we call \textit{sparse representation learning}.

\subsubsection{Neural Weighting Schemes}
One of basic methods to leverage the advantage of neural models while still employing sparse term-based retrieval is to re-weight the term importance before indexing. To this purpose, a direct way is to design neural models to predict term weights based on semantics rather than pre-defined heuristic functions. An alternative method is to augment each document with additional terms, then, expanded documents are stored and indexed with classical term-based methods.

One of the earliest methods to learn term weights is the DeepTR model~\cite{zheng2015learning}, which leverages neural word embeddings to estimate the term importance. Specifically, it constructs a feature vector for each query term and learns a regression model to map feature vectors onto ground truth weights of terms. Estimated weights can be directly used to replace classical term weighting schemes in the inverted index, e.g., BM25 and LM, to generate bag-of-words query representations to improve the retrieval performance.
More recently, \citet{frej2020learning} proposed a term discrimination values (TDVs) learning method, which replaces the IDF field in the original inverted index based on FastText~\cite{bojanowski2017enriching}. In addition to the pairwise ranking objective, they also minimized the $\ell_1$-norm of bag-of-words document representations to reduce the memory footprint of the inverted index and speed up the retrieval process.
Besides, \citet{zuccon2015integrating} used word embeddings within the translation language model for information retrieval. They leveraged word embeddings to estimate translation probabilities between words. This language model captures implicit semantic relations between words in queries and those in relevant documents, thus bridging the vocabulary mismatch and producing more accurate estimations of document relevance.

In recent years, contextual word embeddings, which are often learned with pre-trained language models, have achieved great success in many NLP tasks~\cite{peters2018deep, devlin2018bert, yang2019xlnet}. 
Compared with static word embeddings (e.g., Word2Vec~\cite{mikolov2013Distributed}, GloVe~\cite{pennington2014glove}, and FastText~\cite{bojanowski2017enriching}), contextual word embeddings model the semantic information of words under the global context.
There are also several works trying to utilize contextual word embeddings to estimate term weights. 
For example, \citet{dai2019context, dai2020context} proposed a BERT-based framework (DeepCT) to evaluate the term importance of sentences/passages in a context-aware manner. It maps contextualized representations learned by BERT to term weights, then uses predicted term weights to replace the original TF field in the inverted index. Experimental results show that predicted weights could better estimate the term importance and improve term-based methods for the first-stage retrieval. Moreover, results in~\cite{mackenzie2020efficiency} verify that DeepCT can improve search efficiency via static index pruning technique.
Furthermore, Dai et al.~\cite{dai2020context_www} introduced the HDCT model to learn term weights for long documents. It firstly estimates passage-level term weights using contextual term representations produced by BERT. Then, passage-level term weights are combined into document-level term weights through a weighted sum. 
It is worth noting that the learned term weights by the above models, including DeepCT and HDCT, are in the range of 0-1. Then, they scale the real-valued predictions into a \textit{tf}-like integer. In this way, these term weights can be directly integrated into the existing inverted index and be implemented with existing retrieval models.

Above mentioned approaches rely on neural embeddings, which are learned within local or global contexts, to predict term weights directly. 
Besides, there are also some works trying to estimate term weights by evaluating the matching score between each term and the whole document through a complex interaction network.
For example, \citet{mitra2019incorporating} proposed to incorporate query term independence assumption into three state-of-the-art neural ranking models (BERT~\cite{devlin2018bert}, Duet~\cite{mitra2017learning}, and Conv-KNRM~\cite{dai2018convolutional}), and the final relevance score of the document can be decomposed with respect to each query term. In this way, these neural ranking models can be used to predict the matching score of each term to the document, which can be pre-computed and indexed offline. Experimental results on a passage retrieval task show that this method exhibits significant improvement over classical term-based methods, with only a small degradation compared with original neural ranking models.
Similarity, \citet{mitra2020conformer} extended the Transformer-Kernel~\cite{hofstatter2020interpretable} architecture to the full retrieval setting by incorporating the query term independence assumption. Firstly, they simplified the query encoder by getting rid of all Transformer layers and only considering non-contextualized embeddings for query terms. Secondly, instead of applying the aggregation function over the full interaction matrix, they applied it to each row of the matrix individually, which corresponds to an individual matching score between each query term and the whole document.

In addition to explicitly predicting term weights, another kind of method is to augment the document with additional terms using neural sequence-to-sequence (seq2seq) models. In this way,  term weights of those elite terms can be promoted in the inverted index. 
In fact, this kind of method follows the idea of document expansion described in Section \ref{Document Expansion}, yet it does the expansion with neural networks.
For example, the doc2query~\cite{nogueira2019document} model trains a seq2seq model based on relevant query-document pairs. Then, the seq2seq model generates several queries for each document, and those synthetic queries are appended to the original document, forming the ``expanded document''. This expansion procedure is performed on every document in the corpus, and the expanded document collection is indexed as usual. Finally, it relies on a BM25 algorithm to retrieve relevant candidates. When combined with a re-ranking component, it achieves the state-of-the-art performance on MS MARCO~\cite{nguyen2016msmarco} and TREC CAR~\cite{dietz2017trec} retrieval benchmarks. 
Later, the docTTTTTquery model~\cite{nogueira2019doc2query} employs a stronger pre-trained model T5~\cite{raffel2019exploring} to generate queries and achieves large gains compared with doc2query.
Moreover, \citet{yan2021unified} proposed a Unified Encoder-Decoder networks (UED) to enhance document expansion with the document ranking task. Experimental results on two large-scale datasets show that UED achieves a new state-of-the-art performance on both MS MARCO passage retrieval task and TREC 2019 Deep Learning Track.

There are also some works trying to learn term weights as well as document expansion simultaneously in a unified framework~\cite{bai2020sparterm, mallia2021learning, formal2021splade}.
For example, \citet{bai2020sparterm} proposed a novel framework SparTerm to build term-based sparse representations in the full vocabulary space. It takes the pre-trained language model to map the frequency-based BoW representation to a sparse term importance distribution in the whole vocabulary. In this way, it can simultaneously learn the weights of existing terms and expand new terms for the document. Besides, SparTerm also constructs a gating controller to generate binary and sparse signals across the dimension of vocabulary size, ensuring the sparsity of final representations. 
Besides, DeepImpact~\cite{mallia2021learning} leverages docTTTTTquery to enrich the document collection, and then uses a contextualized language model to estimate the semantic importance of tokens in the document. In this way, it can produce a single-value representation for the original token and expanded token in each document.

\subsubsection{Sparse Representation Learning}
In contrast to weighting terms in the symbolic space, sparse representation learning methods focus on building sparse vectors for queries and documents, where representations are expected to capture semantic meanings of each input text. In this way, queries and documents are represented in the latent space. But different from topic models in Section \ref{Topic Models}, each dimension of the latent space learned by neural models has no clear concepts.
Then, the learned sparse representations can be stored and searched with an inverted index efficiently, where each unit in the inverted index table corresponds to a ``latent word'' instead of a term.

Learning sparse embeddings can be traced back to semantic hashing~\cite{salakhutdinov2009semantic}, which employs deep auto-encoders for semantic modeling. 
It takes a multi-layer auto-encoder to learn distributed representations for documents. 
This model captures the document-term information, but it does not model the relevance relationship between queries and documents.
Thus, it still cannot outperform classical term-based retrieval models, such as BM25 and QL.
\citeauthor{zamani2018neural}~\cite{zamani2018neural} proposed a standalone neural ranking model to learn latent sparse representation for each query and document. Specifically, it firstly maps each n-gram in queries and documents to a low-dimensional dense vector to compress the information and learn the low dimensional manifold of the data. Then, it learns a function to transform n-gram representations to high-dimensional sparse vectors. Finally, the dot product is used as the matching function to calculate the similarity between each query and document. This architecture learns latent sparse representations to better capture semantic relationships between query-document pairs, showing better performance over traditional term-based retrieval and several neural ranking models. But it uses n-gram as an encoding unit, which can only capture local dependencies and cannot adjust dynamically to the global context.
Recently, \citet{jang2021uhd} presented UHD-BERT, a novel sparse retrieval method empowered by extremely high dimensionality and controllable sparsity. They showed that the model outperforms previous sparse retrieval models significantly and delivers competitive performance compared to dense retrieval models.

To make interaction-focused models applicable for the first-stage retrieval, \citet{ji2019efficient} proposed to use sparse representations to improve the efficiency of three interaction-focused neural algorithms (DRMM~\cite{guo2016deep}, KNRM~\cite{xiong2017end}, and Conv-KNRM~\cite{dai2018convolutional}). The work investigates a Locality Sensitive Hashing (LSH~\cite{datar2004locality}) approximation of three neural methods with fast histogram-based kernel calculation and term vector precomputing for a runtime cache. Evaluation results show that the proposed method yields 4.12x, 80.54x, and 106.52x time speedups for DRMM, KNRM, and Conv-KNRM respectively on the ClueWeb dataset.

\begin{figure}[!t]
\centering
\includegraphics[scale=0.85]{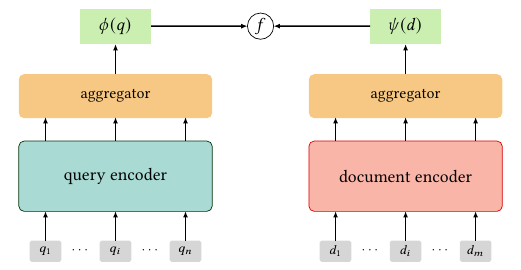}
\caption{Dual-encoder architecture of dense retrieval methods.} 
\label{fig:dual_encoder}                                  
\end{figure}

\subsection{Dense Retrieval Methods}
One of the biggest benefits of neural retrieval methods is to move away from sparse representations to dense representations, which is able to capture semantic meanings of input texts for better relevance evaluation.
As shown in Figure \ref{fig:dual_encoder}, dense retrieval models usually have dual-encoder architecture, also called Siamese network~\cite{bromley1994signature}, which consists of twin networks that accept distinct inputs (queries and documents) and learn standalone dense embeddings for them independently. 
Then, the learned dense representations $\phi(q)$ and $\psi(d)$ are fed into a matching layer $f$, which is often implemented via a simple similarity function, to produce the final relevance score. To support the online serving, the learned dense representations are often indexed and searched via approximate nearest neighbor (ANN) algorithms~\cite{chen2018sptag, johnson2019billion}.

Researchers have devoted a lot of effort to designing sophisticated architectures to learn dense representations for retrieval. Due to the heterogeneous nature of text retrieval, the document often has abundant contents and complicated structures, so that much attention has been paid to the design of the document-side representation function $\psi$. According to the form of learned document representations, we can divide dense retrieval models into two classes, as shown in Figure \ref{fig:dense_retrieval}, \textit{term-level representation learning} and \textit{document-level representation learning}.

\subsubsection{Term-level Representation Learning}
\begin{figure}[!t]
\centering
\includegraphics[scale=0.9]{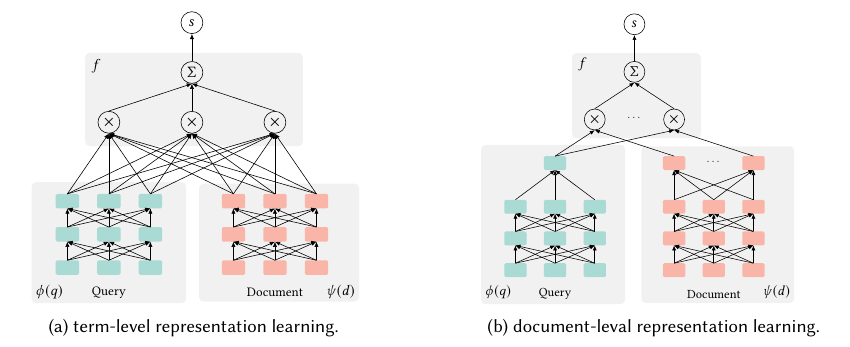}
\caption{Different dense retrieval models for the first-stage retrieval.} 
\label{fig:dense_retrieval}                                  
\end{figure}

Term-level representation learning methods learn fine-grained term-level representations for queries and documents, and queries and documents are represented as a sequence/set of term embeddings. As is shown in Figure \ref{fig:dense_retrieval} (a), the similarity function $f$ then calculates term-level matching scores between the query and the document and aggregates them as the final relevance score.

One of the easiest methods is to take word embeddings, which have been proved to be effective in building ranking models for later re-ranking stages \cite{guo2016deep, xiong2017end}, to build term-level representations for queries and documents.
For example, \citet{Kenter2015Short} investigated whether it is possible to rely only on semantic features, e.g., word embeddings, rather than syntactic representations to calculate similarities between short texts. They replaced the $tf(q_i, d)$ in BM25 with the maximum cosine similarity between the word embedding of $q_i$ and words in the document $d$. Their results show that the model can outperform baseline methods that work under the same condition.
Mitra et al.~\cite{mitra2016dual} trained a word2vec embedding model on a large unlabelled query corpus, but in contrast to only retain the output lookup table, they retained both input and output projections, allowing to leverage both embedding spaces to derive richer distributional relationships. During ranking, they mapped query words into the input space and document words into the output space, and computed the relevance score by aggregating cosine similarities across all query-document word pairs. The experimental results show that the DESM can re-rank top documents returned by a commercial Web search engine, like Bing, better than a term-based signal like TF-IDF. However, when retrieving in a non-telescoping setting, DESM features are very susceptible to false positive matches and can only be used either in conjunction with other document ranking features, such as TF-IDF, or for re-ranking a smaller set of candidate documents.

\begin{figure}[!t]
\centering
\includegraphics[scale=0.95]{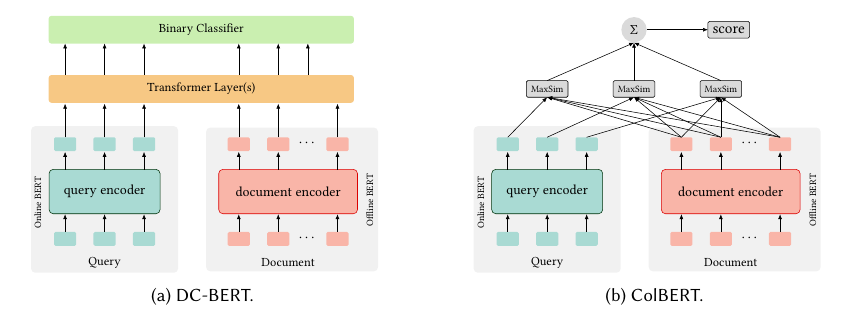}
\caption{Term-level representation learning methods. The left sub-figure shows the architecture of DC-BERT~\cite{zhang2020dc} model, and the right shows the architecture of ColBERT~\cite{khattab2020colbert}.} 
\label{fig:BERT_var}                                  
\end{figure}

\begin{figure}[!t]
\centering
\includegraphics[scale=1.0]{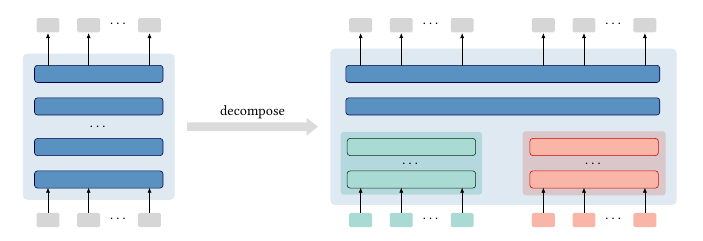}
\caption{Decompose BERT to question-wide and passage-wide self-attentions.} 
\label{fig:BERT_decompose}                                  
\end{figure}

In recent years, the combination of contextual word embeddings and self-supervised pre-training has revolutionized the field of NLP and obtained state-of-the-art performance on many NLP tasks~\cite{peters2018deep, devlin2018bert, yang2019xlnet}.
There are also a number of works that employ contextual word embeddings to learn query/document representations for IR.
For example, \citet{zhang2020dc} proposed the DC-BERT which employs dual BERT encoders for low layers, as shown in Figure \ref{fig:BERT_var} (a), where an online BERT encodes the query only once and an offline BERT pre-encodes all documents and caches all term representations. Then, the obtained contextual term representations are fed into high-layer Transformer interaction, which is initialized by the last $k$ layers of the pre-trained BERT~\cite{devlin2018bert}. The number of Transformer layers $K$ is configurable to a trade-off between the model capacity and efficiency. On the SQuAD dataset and Natural Questions dataset, DC-BERT achieves 10x speedup over the original BERT model on document retrieval, while retaining most (about 98\%) of the QA performance compared to state-of-the-art approaches for open-domain question answering. 
An alternative way to use BERT for the term-level representation learning is the ColBERT \cite{khattab2020colbert} model, which employs a cheap yet powerful interaction function, i.e., a term-based MaxSim, to model fine-grained matching signals, as shown in Figure \ref{fig:BERT_var} (b). Concretely, every query term embedding interacts with all document term embeddings via a MaxSim operator, which computes maximum similarity (e.g., cosine similarity or L2 distance), and scalar outputs of these operators are summed across query terms. Based on this, it can achieve cheap interaction and high-efficient pruning for top-$k$ relevant documents retrieval. Results on MS MARCO and TREC CAR show that ColBERT’s effectiveness is competitive with existing BERT-based models (and outperforms every non-BERT baseline), while executing two orders-of-magnitude faster and requiring four orders-of-magnitude fewer FLOPs per query.
A similar model COIL is proposed by~\citet{gao2021coil}, but the query term embedding only interacts with exactly matched document term embeddings in the MaxSim operator. Experimental results show that COIL performs on par with more expensive and complex all-to-all matching retrievers (e.g., ColBERT).
Besides, \citet{cao2020deformer} and \citet{macavaney2020efficient} proposed DeFormer and PreTTR to decompose lower layers of BERT, which substitutes the full self-attention with question-wide and passage-wide self-attentions, as shown in Figure \ref{fig:BERT_decompose}. The proposed approaches considerably reduce the query-time latency of deep Transformer networks. The difference between them is that the PreTTR model~\cite{macavaney2020efficient} inserts a compression layer to match attention scores to reduce the storage requirement up to 95\% but without substantial degradation in retrieval performance.

A natural extension of the term-level representation learning is to learn phrase-level (i.e, n-grams, sentences) representations for documents, and documents are finally represented as a sequence/set of embeddings. Meanwhile, the query is usually viewed as one phrase and abstracted into a single vector as it is often short in length. Then, the similarity function $f$ calculates matching scores between the query with all phrases in the document and aggregates these local matching signals to produce the final relevance score.
For example, \citet{seo2018phrase} proposed to learn phrase representations based on BiLSTM for OpenQA task. It leads to a significant scalability advantage since encodings of answer candidate phrases in the document can be pre-computed and indexed offline for efficient retrieval. Subsequently, \citet{seo2019real} and \citet{lee2020contextualized} replaced the LSTM-based architecture with a BERT-based encoder, and augmented dense representations learned by BERT with contextualized sparse representations, improving the quality of each phrase embedding. Different from the document encoder, the query encoder only generates one embedding in capturing the whole contextual information of queries. Experimental results show that the OpenQA model that augments learned dense representations with learned contextual sparse representations outperforms previous OpenQA models, including recent BERT-based pipeline models, with two orders of magnitude faster inference time.
For the multi-hop OpenQA task, \citet{feldman2019multi} proposed the MUPPET model for efficient retrieval. The retrieval is performed by considering similarities between the question and contextualized sentence-level representations of the paragraph in the knowledge source. Given the sentence representations $(\bm{s_1}, \bm{s_2} , \dots, \bm{s_k} )$ of a paragraph $P$, and the question encoding $\bm{q}$ for $Q$, the relevance score of $P$ with respect to a question $Q$ is calculated in the following way: 
\begin{equation}
\operatorname{s}(Q, P)=\max _{i=1, \ldots, k} \sigma\left(\left[\begin{array}{c}\bm{s}_{i} \\ \bm{s}_{i} \odot \bm{q} \\ \bm{s}_{i} \cdot \bm{q} \\ \bm{q}\end{array}\right] \cdot\left[\begin{array}{c}\bm{w}_{1} \\ \bm{w}_{2} \\ w_{3} \\ \bm{w}_{4}\end{array}\right]+b\right),
\end{equation}
where $\bm{w_1}, \bm{w_2}, \bm{w_4} \in \mathbb{R}^d$ and $w_3, b \in \mathbb{R}$ are learned parameters. The method achieves state-of-the-art performance over two well-known datasets, SQuAD-Open and HotpotQA.

\subsubsection{Document-level Representation Learning}
The document-level representation learning methods learn one or more coarse-level global representation(s) for each query and each document by abstracting their semantic meanings with dense vectors. It often employs a simple similarity function $f$ (e.g., dot product, or cosine similarity) to calculate the final relevance score based on the query embedding $\phi(q)$ and the document embedding(s) $\psi(d)$, as is shown in Figure \ref{fig:dense_retrieval} (b).

Initial attempts to obtain query embeddings and document embeddings are to directly aggregate their corresponding word embeddings with some pre-defined heuristic functions.
\citet{clinchant2013aggregating} was the first to propose a document representation model, Fisher Vector (FV), based on continuous word embeddings. It firstly maps word embeddings into a higher-dimensional space, then aggregates them into a document-level representation through the fisher kernel framework. Although the FV model outperforms latent semantic indexing (LSI) for ad-hoc retrieval tasks, it does not perform better than classical IR models, such as TF-IDF and the divergence from randomness~\cite{Amati2002Probabilistic} retrieval model. 
\citet{gillick2018end} proposed to utilize the average of word embeddings as the query or document representation. The experimental results show the proposed model outperforms term-based retrieval models (e.g., TF-IDF and BM25), which indicates dense retrieval is a viable alternative to the discrete retrieval model. 
Obtaining text representations by aggregating word embeddings loses the contextual and word orders information as classical term-based retrieval models do. To solve this problem, \citet{le2014distributed} proposed Paragraph Vector (PV), an unsupervised algorithm that learns fixed-length representations from variable-length pieces of texts, such as sentences, paragraphs and documents. ~\citet{ai2016improving, ai2016analysis} evaluated the effectiveness of PV representations for ad-hoc retrieval, but produced unstable performance and limited improvements.
With many attempts that use word/document embeddings to obtain dense representations for queries and documents, only moderate and local improvements over traditional term-based retrieval models have been observed, suggesting the need for more IR-customized embeddings or more powerful representation learning models.

As for embeddings customized for IR, \citet{ai2016analysis} analyzed intrinsic problems of the original PV model that restrict its performance on retrieval tasks. Then, they produced modifications to the PV model, making it more suitable for IR tasks. The evaluation results on Robust04 and GOV2 show the effectiveness of the enhanced PV model.
Subsequently, \citet{gysel2018neural} proposed the Neural Vector Space Model (NVSM), an unsupervised method that learns latent representations of words and documents from scratch for news article retrieval. The query is represented by averaging its constituent word representations and projected to the document feature space. The matching score between a document and a query is given by the cosine similarity between their representations in document feature space. The experiments show that the NVSM outperforms lexical retrieval models on four article retrieval benchmarks.
Similar to NVSM, another unsupervised embedding learning method tailored for IR is SAFIR~\cite{agosti2020learning}. SAFIR jointly learns word, concept and document representations from scratch. The similarity of a query to a document is calculated by averaging its word-concept representations and then projecting it into the document space. Finally, the matching score between the query and the document is given by the cosine similarity between their representations in the document space. The evaluation on shared test collections for medical literature retrieval shows the effectiveness of SAFIR in terms of retrieving relevant documents.
In addition to optimizing word/document embeddings for retrieval objectives directly, considering external knowledge resources, e.g, semantic graphs, ontologies and knowledge graphs, to enhance embeddings learning for semantic retrieval is another effective solution~\cite{nguyen2017learning, tamine2019offline, liu2016constraining}. For example, \citet{liu2016constraining} leveraged the existing knowledge (word relations) in the medical domain to constrain word embeddings using the principle that related words should have similar embeddings. The resulting constrained word embeddings are used for IR tasks, showing superior effectiveness to unsupervised word embeddings.

For more powerful representation learning models for the first-stage retrieval, \citet{henderson2017efficient} proposed a computationally efficient neural method for natural language response suggestion. The feed-forward neural network uses n-gram embedding features to encode messages and suggested replies into vectors, which is optimized to give message-response pairs higher dot product values.
The DPR~\cite{karpukhin2020dense} model is proposed to learn dense embeddings for text blocks with a BERT-based dual encoder. The retriever based on the DPR model outperforms a strong Lucene BM25 system on a wide range of OpenQA datasets and is beneficial for the end-to-end QA performance. 
Similar to DPR, the RepBERT~\cite{zhan2020repBERT} model employs a dual encoder based on BERT to obtain query and document representations, then inner products of query and document representations are regarded as relevance scores. Experimental results show that the RepBERT outperforms BM25 on the MS MARCO passage ranking task.

Another alternative approach is to distill a more complex model (e.g., term-level representation learning method or interaction-focused model) to a document-level representation learning architecture.
For example, \citet{lin2020distilling} distilled the knowledge from ColBERT’s expressive MaxSim operator for computing relevance scores into a simple dot product, thus enabling a single-step ANN search. Their key insight is that during distillation, tight coupling between the teacher model and the student model enables more flexible distillation strategies and yields better learned representations. The approach improves query latency and greatly reduces the onerous storage requirement of ColBERT, while only making modest sacrifices in terms of effectiveness. 
\citet{tahami2020distilling} utilized knowledge distillation to compress the complex BERT cross-encoder network as a teacher model into the student BERT bi-encoder model. This increases the prediction quality of BERT-based bi-encoders without affecting its inference speed. They evaluated the approach on three domain-popular datasets, and results show that the proposed method achieves statistically significant gains.

It should be noted that among neural models proposed early for IR tasks, such as DSSM~\cite{huang2013learning}, ARC-I~\cite{hu2014convolutional} and QA\_LSTM~\cite{tan2015lstm}, they learn highly abstract document representations based on different network architectures, such as fully connection, CNN, and RNN. Then a simple matching function, such as cosine similarity and bilinear, is used to evaluate similarity scores. These models are usually proposed for re-ranking stages at the beginning, however, because of their dual-encoder architecture,  it is theoretically that they are also applicable for the first-stage retrieval. Nevertheless, a study by \citet{guo2016deep} shows that DSSM, C-DSSM~\cite{shen2014learning}, and ARC-I perform worse when trained on a whole document than when trained only on titles. Due to these limitations, most of these early neural models fail to beat unsupervised term-based retrieval baselines (e.g., BM25) on academic benchmarks. These drawbacks motivate the development of models discussed in this survey that are designed specifically for the retrieval stage.

\begin{figure}[!t]
\centering
\includegraphics[scale=0.9]{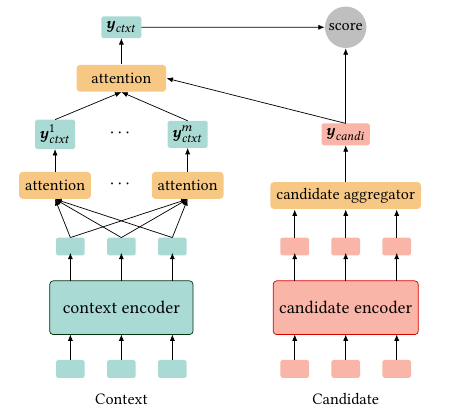}
\caption{Document-level multi-vector representation method in Poly-encoders~\cite{humeau2019poly}.} 
\label{fig:dense_poly}                                  
\end{figure}

In addition to learn a single global representation for each query and each document, another more sophisticated approach is to employ different encoders for queries and documents, where the document encoder abstracts the content into multiple embeddings---each embedding captures some aspects of the document, while the query encoder obtains a single embedding for each query~\cite{luan2020sparse, humeau2019poly, tang2021improving}. The motivation is that documents are often lengthy and have diverse aspects in them, but queries are usually short and have focused topics.
For example, \citet{luan2020sparse} proposed the Multi-Vector BERT (ME-BERT) to obtain a single-vector representation for the query and a multi-vector representation for the document. They represented the sequence of contextualized query/document embeddings at the top level of a deep Transformer, then defined the single-vector query representation as the contextualized embedding of the special token ``\texttt{[CLS]}'' and the multi-vector document representation as the first $m$ contextualized vectors of tokens in the document. The value of $m$ is always smaller than $N$, where $N$ is the number of tokens in the document. Finally, the relevance score is calculated as the largest inner product yielded by each document vector with the query vector. Experimental results show that the ME-BERT model yields strong performance than alternatives in open retrieval. 
Similarly, \citet{humeau2019poly} proposed the Poly-encoders, an architecture with an additional learned attention mechanism to represent more global features. The Poly-encoders, as shown in Figure \ref{fig:dense_poly}, uses two separate Transformer models to encode contexts and candidates. 
The candidate is encoded into a single vector $\bm{y}_{\text{candi}}$, and the input context, which usually includes more information than a candidate, is represented with $m$ vectors $(\bm{y}_{\text{ctxt}}^1, \dots, \bm{y}_{\text{ctxt}}^m)$ instead of just one. Then, $m$ vectors are attended using the candidate encoding vector $\bm{y}_{\text{candi}}$ to get the final score. The value of $m$ will give a trade-off between inference accuracy and speed. 
It should be noted that different from general retrieval tasks that retrieved texts (documents) are usually longer than input texts (queries), the task in \cite{humeau2019poly} has longer input texts than retrieved texts, thus, the multi-vector representation model is actually employed for the query encoder in \cite{humeau2019poly}.

\subsection{Hybrid Retrieval Methods}
\begin{figure}[!t]
\centering
\includegraphics[scale=0.95]{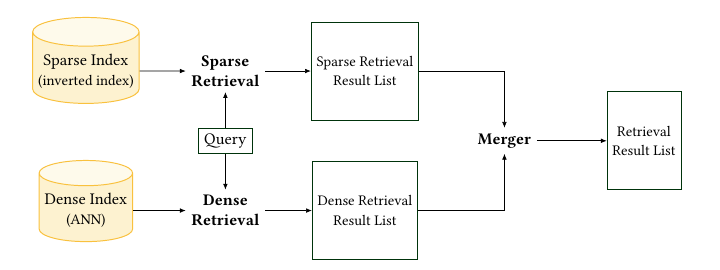}
\caption{The general architecture of hybrid retrieval methods.} 
\label{fig:hybrid_model}                                  
\end{figure}

Sparse retrieval methods take words or ``latent words'' as the unit of indexing, which preserves strong discriminative power as the score is calculated by hard matching between each unit. 
As a result, they can identify exact matching signals, which are momentous for retrieval tasks. On the other hand, dense retrieval methods learn continuous embeddings to encode semantic information and soft matching signals, but detailed low-level features are always sacrificed. A natural approach to balance between the fidelity of sparse retrieval methods and the generalization of dense retrieval methods is to combine merits of them to build a hybrid retrieval model~\cite{vulic2015monolingual, gysel2018neural, gao2020complementing, kuzi2020leveraging}.
Hybrid retrieval methods define multiple representation functions ($\phi$ and $\psi$), and then obtain sparse and dense representations for queries/documents. Finally, these representations are used to calculate the final matching score with different merging ways ($f$). The general architecture of hybrid retrieval methods is shown in Figure~\ref{fig:hybrid_model}.

With the development of the word embedding technique, there are a number of works on exploiting it with term-based models for the first-stage retrieval.
\citet{vulic2015monolingual} obtained better results on monolingual and bilingual retrieval by combining the word-embedding-based method with a uni-gram language model.
However, the embedding-based model solely does not outperform traditional language models in the monolingual retrieval task. That is to say, the effectiveness of neural semantic retrieval models is more observed when combined with term-based retrieval methods, instead of replacing them.
The consistent observation is also obtained in~\cite{mitra2016dual, Nalisnick2016Improving}, that direct using of word embeddings only obtains extremely poor performance in the non-telescoping setting, unless combining it with a term-based feature, such as BM25.
The GLM~\cite{ganguly2015word} is an embedding-based translation model linearly combined with a traditional language model. The probability of observing a term $t$ in the query from a document $d$ is modeled by three parts, i.e., direct term sampling, generating a different term $t'$ either from the document itself or from the collection, and then transforming it to the observed query term $t$. The empirical results show that GLM performs better than the traditional language model.
\citet{roy2016representing} also proposed to combine word vector based query likelihood with the standard language model based query likelihood for document retrieval. Experiments on standard text collections show that the combined similarity measure almost always outperforms the language model similarity measure significantly.
Besides, according to the experimental results got in ~\cite{gysel2018neural}, although the NVSM model outperforms term-based retrieval models on some benchmarks, it will be more useful as a supplementary signal to term-based models. Similar conclusions could also be found in~\cite{agosti2020learning, tamine2019offline, liu2016constraining}.

\begin{figure}[!t]
\centering
\includegraphics[scale=1.0]{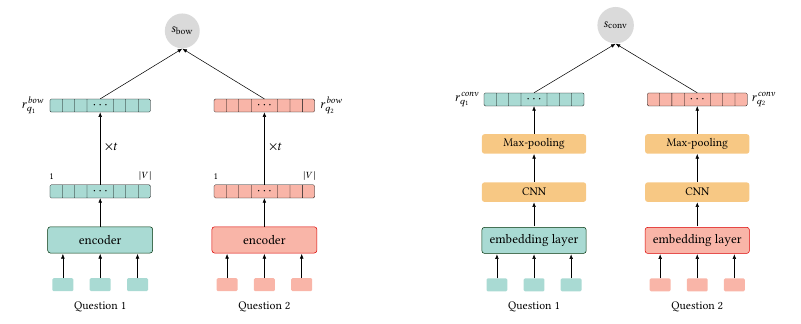}
\caption{The sparse and dense scoring component in BOW-CNN~\cite{dos-santos-etal-2015-learning}.} 
\label{fig:dense_sparse}                                  
\end{figure}

Different from using word embeddings to construct dense representations and using term frequency to obtain term-based matching scores directly, there are also some works trying to employ simple neural networks to learn sparse and dense representations, then calculate matching scores based on learned representations.  
For example, \citet{dos-santos-etal-2015-learning} proposed the BOW-CNN architecture to retrieve similar questions in online QA community sites, as shown in Figure~\ref{fig:dense_sparse}, which combines a bag-of-words (BOW) representation with a distributed vector representation created by a convolutional neural network (CNN). 
The BOW-CNN model computes two partial similarity scores: $s_{\mathrm{bow}}(q_1, q_2)$ for BOW representations and $s_{\mathrm{conv}}(q_1, q_2)$ for CNN representations. Finally, it combines  two partial scores to create the final score $s(q_1, q_2)$. They performed experiments on two datasets collected from Stack Exchange communities. The experimental results evidence that BOW-CNN is more effective than BOW-based information retrieval methods such as TF-IDF, and BOW-CNN is more robust than the pure CNN for long texts.
Besides, \citet{MacAvaney2020Expansion} proposed a new approach for passage retrieval, which trains a model to generate query and document representations in a given fixed-length vector space, and produce a ranking score by computing a similarity score between two representations. Different from other representation learning methods, it represents each query as a sparse vector and each document as a dense vector. Finally, the dot product is used to compute the similarity between the query vector and the document vector. The experimental results show that the proposed EPIC model significantly outperforms prior approaches. It is also observed that the performance is additive with current leading first-stage retrieval methods.

With the rise of more powerful pre-training neural networks (e.g., BERT, GPT-3), it is a natural way to combine them with term-based models for improving the first-stage retrieval.
\citet{seo2019real} proposed the DenSPI for the retrieval stage of OpenQA. The DenseSPI model constructs the dense-sparse representation for each phrase unit. The dense vector is represented as pointers to the start and end BERT-based token representations of the phrase, which is responsible for encoding syntactic or semantic information of the phrase with respect to its context. The sparse embedding uses 2-gram-based tf-idf for each phrase, which is good at encoding precise lexical information.
Later, \citet{lee2020contextualized} proposed to learn contextual sparse representation for each phrase based on BERT to replace term-frequency-based sparse encodings in DenSPI~\cite{seo2019real}. This method leverages rectified self-attention to indirectly learn sparse vectors in n-gram vocabulary space, improving the quality of each phrase embedding by augmenting it with a contextualized sparse representation. Experimental results show that the OpenQA model that augments DenSPI with learned contextual sparse representations outperforms previous OpenQA models, including recent BERT-based pipeline models, with two orders of magnitude faster inference time.
\citet{luan2020sparse} proposed to linearly combine the term-based system (BM25-uni) and neural-based system (dual-encoder or multi-vector model) scores using a single trainable weight $\lambda$, tuned on a development set, which yields strong performance while maintaining the scalability.
\citet{gao2020complementing} proposed the CLEAR model, which uses a BERT-based embedding model to complement the term-based model (BM25). Experimental results show that retrieval from CLEAR without re-ranking is already almost as accurate as the BERT re-ranking pipeline.
Similarly, \citet{kuzi2020leveraging} proposed a general hybrid approach for document retrieval that leverages both a semantic model (BERT) and a lexical retrieval model (BM25). An in-depth empirical analysis is performed, which demonstrates the effectiveness of the hybrid approach and also sheds some light on the complementary nature of the lexical and semantic models.

\subsection{Model Learning}
As described above, neural semantic retrieval models always define functions $\phi$, $\psi$, and $f$ in the network structure. These functions are usually learned from data using deep learning technology. Here, we discuss key topics on the learning of neural semantic retrieval models, including loss functions and negative sampling strategies.

\subsubsection{Loss Functions}
We review major training objectives adopted by neural semantic retrieval models. 
Ideally, after the training loss is minimized, all preference relationships between documents should be satisfied and the model will produce the optimal result list for each query. This makes training objectives effective in many tasks where performance is evaluated based on the ranking of relevant documents.

In practice, the most commonly used loss function is \textit{sampled cross entropy loss}, also called negative log likelihood loss:
\begin{equation}
\mathcal{L}\left(q, d^{+}, D^{-}\right)=-\log \frac{\exp \left(s\left(q, d^{+}\right)\right)}{\exp \left(s\left(q, d^{+}\right)\right)+\sum_{d^{-} \in D^{-}} \exp \left(s\left(q, d^{-}\right)\right)},
\label{eq:nll}
\end{equation}
where $q$ denotes a query, $d^{+}$ is a relevant document of $q$, and $D^{-}$ is the irrelevant document set of $q$.

Another commonly used loss function is the \textit{hinge loss}: 
\begin{equation}
\mathcal{L}\left(q, d^{+}, D^{-}\right)=\frac{1}{n} \sum_{d^{-} \in D^{-}} \max \bigl(0,m-\left(s(q, d^{+})-s(q, d^{-})\right)\bigl),
\label{eq:hinge}
\end{equation}
where $q$ denotes a query, $d^{+}$ is a relevant document of $q$, $D^{-}$ is the irrelevant document set of $q$, $n$ is the number of documents in $D^{-}$, and $m$ is the margin which is usually set as 1.

In fact, the negative log likelihood loss (Eq.~(\ref{eq:nll})) and hinge loss (Eq.~(\ref{eq:hinge})) are also widely used in many other tasks with different names, e.g., InfoNCE loss in contrastive representation learning~\cite{Oord:arxiv18:Representation,chen2020simple} and bayesian personalized ranking (BPR) loss in recommender systems~\cite{Rendle:uai09:BPR}. 
These loss functions and their variations have been well studied in other fields, including extreme multi-class classification~\cite{Blanc:icml18:Adaptive,Rawat:nips19:Sampled,Bamler2020Extreme}, representation learning~\cite{Oord:arxiv18:Representation,chen2020simple}, deep metric learning~\cite{Wang:mm17:NormFace,Wang:spl18:Additive,Sun:cvpr20:Circle}, etc. The research progress in these fields might provide some insights to inspire the loss design in neural semantic retrieval. First of all, \citet{Wang:mm17:NormFace} showed the softmax log likelihood loss is actually a smooth version of hinge loss. Moreover, several works have shown that the concept of margin in hinge loss can also be introduced into softmax cross entropy loss to improve the performance in tasks like face recognition and Person Re-identification~\cite{Wang:spl18:Additive,Sun:cvpr20:Circle}. In addition, works~\cite{Wang:mm17:NormFace,chen2020simple,Yang:www20:Mixed} in different domains all verify that applying the $\ell_2$ normalization to final representations (i.e., using cosine as the score function $f$) along with temperature can make the learning robust and improve the performance. Another line of research focuses on the bias in sampled softmax cross entropy loss~\cite{jean:acl2015:using,chuang2020debiased}. For example, works in NLP~\cite{jean:acl2015:using,Blanc:icml18:Adaptive} usually focus on the unbiased estimation of the full softmax, while \citet{chuang2020debiased} focused on correcting the bias introduced by the false negative samples that have the same label as the ground truth. It is worth noting that these conclusions need to be re-examined under the first-stage retrieval task.

\subsubsection{Negative Sampling Strategies} \label{Negative_Sampling}
In loss functions (Eq.~(\ref{eq:nll}) and Eq.~(\ref{eq:hinge})), the negative example set $D^-$ is an important part of inputs. However, during the learning of the first-stage retrieval models, it is often the case that only positive examples are available in the training dataset, while negative examples are not explicitly labeled. In fact, the sampling strategy of negative examples is a crucial topic in neural semantic retrieval models, because it directly determines the quality of the learned retrieval model.

Negative sampling is a common fundamental problem in the learning of many tasks where only positive signals are explicitly existed, like recommender system~\cite{Zhang:sigir13:Optimizing,Ying:kdd18:Graph,Ding:ijcai19:Reinforced,Yi:recsys19:Sampling,Yang:www20:Mixed}, graph mining~\cite{sun:iclr2018:rotate,Armandpour:aaai19:Robust,Yang:kdd20:Understanding}, and self-supervised representation learning~\cite{Chen:wsdm18:Improving,bose:acl18:adversarial,zhang:acl18:gneg,he2020momentum,chen2020simple}.
Here, we mainly focus on the research progress in the field of the first-stage retrieval.
The neural semantic retrieval models usually vary in their mechanisms to construct negative examples. But in general, negative sampling strategies can be divided into three categories:
\begin{enumerate}[label=(\arabic *)] 
\item \textit{Random Negative Sampling}: random samples from the entire corpus~\cite{luan2020sparse, karpukhin2020dense} or in batch~\cite{zhan2020repBERT, gillick2018end, henderson2017efficient, henderson2019training}.
It should be noted that if using the batch as a source for random negatives, the batch size becomes important~\cite{gillick2018end}. \citet{lee2019latent} suggested to use a large batch size because it makes the training task more difficult and closer to what the retriever observes at test time. However, the batch size is usually restricted by computing resources and cannot be set very largely. To address this problem, \citet{he2020momentum} proposed to decouple the size of mini-batch and sampled negative examples by maintaining a queue of data samples (encoded representations of the current mini-batch are enqueued, and the oldest are dequeued) to provide negative samples. In this way, they can use a very large size (e.g., \num{65536}) for negative samples in unsupervised visual representation learning. 
However, random negative sampling is usually sub-optimal for training neural semantic retrieval models.
Models can hardly focus on improving top ranking performance since these random negative samples are usually too easy to be distinguished.
This problem would lead to serious performance dropping in practice. 
To make the model better at differentiating between similar results, one can use samples that are closer to positive examples in the embedding space as hard negatives for training. 
Thus, mining hard negative samples to optimize retrieval performance is a key problem that needs to be addressed.
\item \textit{Static Hard Negative Sampling}: random samples from pre-retrieved top documents by a traditional retriever~\cite{luan2020sparse, karpukhin2020dense, gao2020complementing}, such as BM25. Recent researches find it helps training convergence to include BM25 negatives to provide stronger contrast for representations learning~\cite{luan2020sparse, karpukhin2020dense}.
Obtaining hard negative samples with pre-retrieval is computationally efficient. 
However, hard negative samples obtained by static methods are not real hard negatives. 
Intuitively, strong negatives close to relevant documents in an effective neural retrieval model space should be different from those from term-based retrieval models, as the goal of neural semantic retrieval models is to find documents beyond those retrieved by term-based models. If using negative samples from BM25, there exists a severe mismatch between negatives used to train the retrieval model and those seen in testing. 
\item \textit{Dynamic Hard Negative Sampling}: random samples from top-ranked irrelevant documents predicted by the retrieval model itself.
Intuitively, negative sampling dynamically according to current semantic retrieval models, e.g., using the distribution which is proportional to relevance scores predicted by the current model, should be a very promising choice for producing informational negative samples~\cite{Park:www2019:Adversarial,sun:iclr2018:rotate}. 
In this way, neural semantic retrieval models can optimize themselves using negative samples they did wrong (i.e., predict a high relevance score for an irrelevant document). 
However, it is usually impractical to score all candidate documents in a very large corpus on the fly. Thus, in real-world settings, periodically refreshing the index and retrieving top-ranked documents as hard negatives is a more practical compromise choice~\cite{xiong2020approximate, huang2020embedding, gao2020complementing, ding2020rocketqa}. For example, hard negatives mining in~\cite{xiong2020approximate} elevates the BERT-based siamese architecture to robustly exceed term-based methods for document retrieval. It also convincingly surpasses concurrent neural semantic retrieval models for passage retrieval on OpenQA benchmarks.
\end{enumerate}

It should be noted that the negative sampling strategies described above are not exclusive mutually.
In practice, random sampled easy negatives and hard negatives are always used simultaneously. For example, the counterintuitive finding in~\cite{huang2020embedding} shows that models trained simply using hard negatives cannot outperform models trained with random negatives. The hypothesis is that the presence of easy negatives in training data is still necessary, as a retrieval model is to operate on an input space which comprises data with mixed levels of hardness, and the majority of documents in the collection are easy cases which do not match the query at all. Having all negatives being such hard will change the representativeness of the training data to the real retrieval task, which might impose a non-trivial bias to learned embeddings.

\vbox{}
\textbf{Takeaway.}
Neural semantic retrieval methods learn the representation functions (i.e, $\phi$ and $\psi$) and the scoring function ($f$) with deep learning technologies. 
To support fast retrieval, document representations are often learned with standalone networks, and pre-computed and stored with delicate structures.
According to how the representations are computed and stored, we summarize neural semantic retrieval methods into three paradigms, i.e., sparse retrieval methods, dense retrieval methods and hybrid retrieval methods.
\begin{itemize}
\item Sparse retrieval methods focus on improving classical term-based methods by either learning to re-weight terms with contextual semantics or mapping texts into ``latent word'' space. Empirical results show that sparse retrieval methods could indeed improve the performance of the first-stage retrieval, and they are easily integrated with the existing inverted index for efficient retrieval. Moreover, these methods often show good interpretability as each dimension of the representation corresponds to a concrete token or a latent word.
\item Dense retrieval methods employ the dual-encoder architecture to learn standalone low-dimensional dense vectors for queries and documents, aiming to capture the global semantics of input texts. To support online services, the learned dense representations are often indexed and searched via approximate nearest neighbor (ANN) algorithms. These methods have shown promising results on several benchmarks (e.g., MS MARCO and TREC CAR), and attracted increasing attention of researchers.
\item Hybrid retrieval methods define multiple representation functions for queries and documents, and then obtain their sparse and dense representations simultaneously for matching. They are able to achieve a balance between the fidelity of sparse retrieval methods and the generalization of dense retrieval methods. As a result, hybrid retrieval methods show better performance in practice, but require much higher space occupation and retrieval complexity.
\end{itemize}
For neural semantic retrieval models learning, the negative sampling strategy is decisive for learning a high-quality retrieval model. Currently, there have been several works to explore better negative sampling methods, but it is still an open problem on how to mine negative documents for efficient and effective model learning.

\section{Challenges and future directions}
In this section, we discuss some open challenges and several future directions related to semantic models for the first-stage retrieval. Some of these topics are important but have not been well addressed in this field, while some are very promising directions for future researches.

\subsection{Pre-training Objectives for the Retrieval Stage}
Starting 2018, there is rapid progress in different NLP tasks with the development of large pre-training models, such as BERT~\cite{devlin2018bert} and GPT~\cite{radford2018improving}. They are pre-trained on the large-scale corpus and general-purpose modeling tasks such that the knowledge can be transferred into a variety of downstream tasks. 
With this intriguing property, one would expect to repeat these successes for IR tasks.

Some researchers~\cite{lee2019latent, guu2020realm, chang2020pre} have explored pre-training models for the retrieval stage with a dual-encoder architecture. For example, \citet{lee2019latent} proposed to pre-train the two-tower Transformer encoder model with the Inverse Cloze Task (ICT) to replace BM25 in the passage retrieval stage for the OpenQA task. The advantage is that the retriever can be trained jointly with the reader. Nevertheless, the pre-training model does not outperform BM25 on the SQuAD dataset, potentially because the fine-tuning is only performed on the query-tower. 
Except for the ICT pre-training task, \citet{chang2020pre} also proposed the Body First Selection (BFS) and Wiki Link Prediction (WLP) tasks, and studied how various pre-training tasks help the large-scale retrieval problem, e.g., passage retrieval for OpenQA. 
The experimental results show that with properly designed paragraph-level pre-training tasks including ICT, BFS, and WLP, the two-tower Transformer encoder model can considerably improve over the widely used BM25 algorithm.
Besides, \citet{ma2020prop, ma2021b} proposed pre-training with the Representative Words Prediction (ROP) task for ad-hoc retrieval, which achieves significant improvement over baselines without pre-training or with other pre-training methods. 
However, whether the ROP task works for the retrieval stage needs to be re-examined since their experiments are conducted under re-ranking stages.

In summary, there has been little effort to design large pre-training models towards the first-stage retrieval task. As is known to all, the first-stage retrieval mainly focuses on the capability to recall potentially relevant documents as many as possible.
Thus, considering retrieval requirements in recalling relevant documents and modeling task-dependent characteristics would be important elements during designing novel pre-training objectives for the retrieval stage. 
Besides, using cross-modal data (e.g., images) to enhance language understanding is also a promising direction in pre-training researches.

\subsection{More Effective Learning Strategies}
For information retrieval tasks, the construction of benchmark datasets often relies on a pooling process to recall a subset of documents for expert judging. Such labeling process leads to the well-known bias problem, where the dataset only contains partially positive documents and the rest of unlabeled documents are oftentimes assumed to be equally irrelevant~\cite{zhan2020repBERT, karpukhin2020dense}. To address the bias problem, it is necessary to devise smart learning strategies to achieve effective and efficient model training. For example, \citet{chuang2020debiased} developed a debiased contrastive objective that corrects for the sampling of the same label data-points, even without knowledge of true labels.
Next, as discussed in Section \ref{Negative_Sampling}, hard negative samples can improve the model’s ability to differentiate between similar examples. However, hard negatives mining strategies have not been fully explored. One of the state-of-the-art methods is the Asynchronous ANCE training proposed by \citet{xiong2020approximate}, which periodically refreshes the ANN index and samples top-ranked documents as negatives. Although ANCE is competitive in terms of effectiveness, refreshing the index periodically greatly increases the model training cost (e.g., 10h for each period).
Besides, some works conclude that it would be more effective to learn semantic retrieval models with hard negative samples and easy negative samples simultaneously~\cite{zhan2021optimizing}. Thus, in addition to mining hard negatives, it is also worthy to explore arranging the position and order of training samples since negative documents often show varied-level of difficulties. We believe it would be interesting and valuable to study more complex training strategies, such as curriculum learning~\cite{bengio2009curriculum}, to help the model optimization for the first-stage retrieval. 
Moreover, the supervised data for IR is always scarce since it requires much manual labor to obtain. Besides, the supervised dataset is prone to long-tail, sparsity and other issues. Thus, weak supervised or unsupervised learning, e.g., contrastive learning~\cite{he2020momentum, chen2020simple}, are promising directions. For example, ~\citet{dai2020context_www} proposed a content-based weak supervision strategy that exploits the internal structure of documents to mine training labels.

\subsection{Benchmark Testbed for Efficiency Comparison}
The multi-stage retrieval paradigm aims to balance between the effectiveness and efficiency of retrieval tasks, where the first-stage retrieval focus on the efficiency and re-ranking stages pay more attention to the effectiveness.
But efficiency metrics in isolation are meaningless unless contextualized with corresponding effectiveness measures. 
Ideally, the efficiency metrics at different effectiveness cutoffs should be reported on the leaderboard.
Moreover, since the customized hardware, e.g., GPUs or TPUs, has a significant impact on the computation time of deep models, and the response time of the first-stage retrieval models is also infamously sensitive to constraints, such as locality of data on file systems for caching, it is expected to compare different models under the same conditions. However, fair conditions for model efficiency comparison have not been fully valued and studied in the IR field as in the computer vision (CV) community. 
For example, the medical computer vision community has already recognized the need for a focus on run time considerations. The medical image analysis benchmark VISCERAL~\cite{Cloud-Based} includes run-time measurements of participant solutions on the same hardware. Additionally, CV tasks, such as object detection and tracking, often require real-time results~\cite{huang2017speed}. 
For IR tasks, \citet{hofstatter2019let} put forward a preliminary solution, which makes the comparison of run time metrics feasible by introducing docker-based submissions of complete retrieval systems so that all systems can be compared under the same hardware conditions by a third party.

\subsection{Advanced Indexing Schemes}
As described in Section~\ref{index}, for IR tasks, indexing schemes play an important role in determining the way to organize and retrieve large-scale documents. Specially, most dense retrieval methods, which learn dense representations for queries and documents, rely on ANN algorithms to perform efficient vector search for online services~\cite{khattab2020colbert, cai2021discriminative}. 

Existing dense retrieval methods always separate two steps of representation learning and index building. This pattern suffers from a few drawbacks in practical scenarios. Firstly, the indexing process cannot benefit from supervised information because it uses the task-independent function to build the index. Besides, the representation and index are separately obtained and thus may not be optimally compatible. These problems all result in severely decayed retrieval performance. In fact, there have been studies~\cite{yu2018product, zhang2021joint} to explore the joint training of encoders and indexes in the fields of image retrieval and recommendation. 
For information retrieval, it is still in its infancy stage to design joint learning schemes of the first-stage retrieval models and indexing methods, and we believe it would be an interesting and promising direction.

On the other hand, how to design better ANN algorithms that can manage large-scale documents and support efficient and precise retrieval is another important direction. Compared with the brute-force search, the essence of ANN search is to sacrifice part of precision to get higher retrieval efficiency. Generally, there are two kinds of ANN algorithms from the principle of improving retrieval efficiency. One is non-exhaustive ANN search methods~\cite{bernhardsson2018annoy, malkov2018efficient}, and the other is vector compression methods~\cite{ge2013optimized, jegou2010product, indyk1998approximate}. However, each method has its limitations or deficiency, where the non-exhaustive method has a large index size and the compression method has suboptimal performance. Thus, with the booming development of dense retrieval methods, it is urgent to develop more advanced ANN search algorithms to achieve a better balance between the efficiency and effectiveness.

\section{Conclusion}
The purpose of this survey is to summarize the current research status on semantic retrieval models, analyze existing methodologies, and gain some insights for future development.
It includes a brief review of early semantic retrieval methods, a detailed description of recent neural semantic retrieval methods and the connection between them. Specially, we pay attention to neural semantic retrieval methods, and review them from three major paradigms, including sparse retrieval methods, dense retrieval methods and hybrid retrieval methods. We also refer to key topics about neural semantic retrieval models learning, such as loss functions and negative sampling strategies.
In addition, we discuss several challenges and promising directions that are important for future researches. We look forward to working with the community on these issues.

We hope this survey can help researchers who are interested in this direction, and will motivate new ideas by looking at past successes and failures. Semantic retrieval models are part of the broader research field of neural IR, which is a joint domain of deep learning and IR technologies with many opportunities for new researches and applications. We are expecting that, through the effort of the community, significant breakthroughs will be achieved for the first-stage retrieval problem in the near future, similar to those happened in re-ranking stages.

\section{Acknowledgements}
This work was funded by Beijing Academy of Artificial Intelligence (BAAI) under Grants No. BAAI2019ZD0306, the National Natural Science Foundation of China (NSFC) under Grants No. 61902381, 62006218, and 61872338, the Youth Innovation Promotion Association CAS under Grants No. 20144310, and 2021100, the Lenovo-CAS Joint Lab Youth Scientist Project, and the Foundation and Frontier Research Key Program of Chongqing Science and Technology Commission (No. cstc2017jcyjBX0059).

\renewcommand\refname{Reference}
\normalem
\bibliographystyle{ACM-Reference-Format}
\bibliography{main} 

\end{document}